\documentclass[twocolumn]{aastex631}
\pdfoutput=1 
\usepackage[T1]{fontenc}
\usepackage{amsmath,amstext}
\usepackage{amssymb}
\usepackage[T1]{tipa}
\usepackage{apjfonts} 
\usepackage[figure,figure*]{hypcap}
\usepackage{microtype}
\usepackage{tablefootnote}
\usepackage{booktabs}  
\usepackage{longtable}
\usepackage{hyperref}
\usepackage{url}
\usepackage{CJKutf8}
\usepackage{graphicx}
\usepackage{placeins}
\usepackage{xcolor}
\newcommand{\dmeg}{\ensuremath{\mathrm{DM_{ext}}}}
\newcommand{\dmunits}{\ensuremath{\mathrm{pc\,cm^{-3}}}}
\newcommand{\dmism}{\ensuremath{\mathrm{DM_{MW_{ISM}}}}}
\newcommand{\dmcosmic}{\ensuremath{\mathrm{DM_{cosmic}}}}
\newcommand{\dmt}{\ensuremath{\mathrm{DM_{FRB}}}}
\newcommand{\dmmwhalo}{\ensuremath{\mathrm{DM_{MW_{halo}}}}}
\newcommand{\dmtarg}{\ensuremath{\mathrm{DM_{M31,halo}}}}

\newcommand{\dmhost}{\ensuremath{\mathrm{DM_{host}}}}

\newcommand{\dmi}{\ensuremath{\mathrm{DM}_{i}}}
\newcommand{\dmg}{\ensuremath{\mathrm{DM}_{\gamma}}}

\newcommand{\dmcs}{\ensuremath{\overline{\mathrm{DM}}_{\mathrm{cs}}}}
\newcommand{\rvir}{\ensuremath{r_{\mathrm{vir}}}}

\newcommand{\ddmstandardev}{\ensuremath{\delta\mathrm{DM_{STD}}}}

\newcommand{\mocksampleog}{$\rm mock_{3000}$}
\newcommand{\mocksamplebig}{$\rm mock_{20,000}$}

\newcommand{\zfrb}{\ensuremath{z_{\mathrm{FRB}}}}

\newcommand{\mbhalo}{\ensuremath{M_{b,\mathrm{halo}}}}

\newcommand{\mbcosmic}{\ensuremath{M_{b,\mathrm{cosmic}}}}
\newcommand{\mhalo}{\ensuremath{M_{\mathrm{halo}}}}

\newcommand{\rperp}{\ensuremath{R_\perp}}

\newcommand{\mcgm}{\mbhalo}
\newcommand{\mstar}{\ensuremath{M_{*}^{\mathrm{M31}}}}
\newcommand{\ddM}{\ensuremath{\delta\mathrm{DM}}}
\newcommand{\rclose}{\ensuremath{r_{\mathrm{close}}}}
\newcommand{\kclose}{\ensuremath{r_{\mathrm{close}}}}
\newcommand{\fb}{\ensuremath{\Omega_b/\Omega_m}}

\newcommand{\virialradius}{\ensuremath{302\,\mathrm{kpc}}}
\newcommand{\distancetoandromeda}{\ensuremath{752\,\mathrm{kpc}}}
\newcommand{\stellarmass}{\ensuremath{(12 \pm 3) \times 10^{10}\,M_\odot}}
\newcommand{\innerhalorange}{\ensuremath{\approx0}--\ensuremath{151\,\mathrm{kpc}}}
\newcommand{\outterhalorange}{\ensuremath{\approx151}--\ensuremath{302\,\mathrm{kpc}}}

\newcommand{\kboundlower}{\ensuremath{0.5}}
\newcommand{\kboundupper}{\ensuremath{30}}

\newcommand{\valddmlower}{\ensuremath{5.9\text{--}59.6\,\dmunits}}
\newcommand{\valddmupper}{\ensuremath{25.7\text{--}64.6\,\dmunits}}
\newcommand{\valddmlowersnr}{\ensuremath{1.2}}
\newcommand{\valddmuppersnr}{\ensuremath{2.3}}

\newcommand{\numfrbsfullsample}{\ensuremath{3469}}
\newcommand{\numfrbsfullsampleinhalo}{\ensuremath{275}}

\newcommand{\numfrbscleansampleinhalo}{\ensuremath{171}}
\newcommand{\numfrbscontrolsample}{\ensuremath{684}}
\newcommand{\numfrbscontrolsampleinhalo}{\ensuremath{171}}
\newcommand{\numfrbscontrolsampleinhaloinnerbin}{\ensuremath{58}}
\newcommand{\numfrbscontrolsampleinhalouterbin}{\ensuremath{113}}
\newcommand{\numfrbscontrolsampleouthalo}{\ensuremath{684}}
\newcommand{\nummockfrbsinhalosmall}{\ensuremath{182}}

\newcommand{\kclosebestfit}{\ensuremath{9.2}}
\newcommand{\kcloseloweronesigma}{\ensuremath{4.3}}
\newcommand{\kcloseupperonesigma}{\ensuremath{19.2}}
\newcommand{\kcloselowertwosigma}{\ensuremath{1.4}}

\newcommand{\kclosemassest}{\ensuremath{18.6^{+7.9}_{-8.4} \times 10^{10}\,M_\odot}}

\newcommand{\adtestdmismpvalue}{\ensuremath{0.25}}

\newcommand{\adtestdmegpvalue}{\ensuremath{0.16}}

\newcommand{\adtestdmegmockpvalue}{\ensuremath{0.25}}

\newcommand{\kkoname}{k'ni\textipa{P}atn k'l$\left._\mathrm{\smile}\right.$stk'masqt}

\shorttitle{Constraining the mass of the M31 ionized baryon Halo using CHIME/FRB Catalog 2}
\shortauthors{Kahinga et al.}

\graphicspath{{./}}

\submitjournal{ApJ}

\begin{document}
\title{Constraining the mass of the M31 ionized baryon Halo using CHIME/FRB Catalog 2}

\newcommand{\CIERA}{\affiliation{Center for Interdisciplinary Exploration and Research in Astronomy, Northwestern University, 1800 Sherman Avenue, Evanston, IL 60201, USA }}

\newcommand{\NU}{\affiliation{Department of Physics and Astronomy, Northwestern University, Evanston, IL 60208, USA}}

\newcommand{\UDOM}{\affiliation{Department of Physics, College of Natural and Mathematical Sciences, University of Dodoma, Dodoma, P.O. Box 259, Tanzania}}

\newcommand{\Uch}
{\affiliation{Department of Astronomy and Astrophysics, University of Chicago, William Eckhart Research Center, 5640 South Ellis Avenue, Chicago, IL 60637, USA}}

\newcommand{\UCSC}{\affiliation{Department of Astronomy and Astrophysics, University of California, Santa Cruz, 1156 High Street, Santa Cruz, CA 95064, USA}}

\newcommand{\IPMU}{\affiliation{Kavli Institute for the Physics and Mathematics of the Universe (Kavli IPMU), 5-1-5 Kashiwanoha, Kashiwa, 277-8583, Japan}}

\newcommand{\NAOJ}{\affiliation{Division of Science, National Astronomical Observatory of Japan, 2-21-1 Osawa, Mitaka, Tokyo 181-8588, Japan}}

\newcommand{\MU}{\affiliation{Department of Physics, McGill University, 3600 rue University, Montr\'eal, QC H3A 2T8, Canada}}

\newcommand{\Trottier}{\affiliation{Trottier Space Institute, McGill University, 3550 rue University, Montr\'eal, QC H3A 2A7, Canada}}

\newcommand{\CMU}{\affiliation{McWilliams Center for Cosmology \& Astrophysics, Department of Physics, Carnegie Mellon University, Pittsburgh, PA 15213, USA}}

\newcommand{\UVA}
{\affiliation{Anton Pannekoek Institute for Astronomy, University of Amsterdam, Science Park 904, 1098 XH Amsterdam, The Netherlands}}

\newcommand{\WVUPA}
{\affiliation{Department of Physics and Astronomy, West Virginia University, PO Box 6315, Morgantown, WV 26506, USA }}

\newcommand{\ASTRON}
{\affiliation{ASTRON, Netherlands Institute for Radio Astronomy, Oude Hoogeveensedijk 4, 7991 PD Dwingeloo, The Netherlands
}}

\newcommand{\MITK}
{\affiliation{MIT Kavli Institute for Astrophysics and Space Research, Massachusetts Institute of Technology, 77 Massachusetts Ave, Cambridge, MA 02139, USA}}

\newcommand{\CALTECH}
{\affiliation{Cahill Center for Astronomy and Astrophysics, MC 249-17 California Institute of Technology, Pasadena CA 91125, USA}}

\newcommand{\MITP}
{\affiliation{Department of Physics, Massachusetts Institute of Technology, 77 Massachusetts Ave, Cambridge, MA 02139, USA}}

\newcommand{\CCAPS}{\affiliation{Cornell Center for Astrophysics and Planetary Science, Cornell University, Ithaca, NY 14853, USA}}

\newcommand{\DI}
{\affiliation{Dunlap Institute for Astronomy and Astrophysics, 50 St. George Street, University of Toronto, ON M5S 3H4, Canada}}

\newcommand{\DoSS}
{\affiliation{Department of Statistical Sciences, University of Toronto, 700 University Ave., Toronto, ON M5G 1Z5, Canada}}

\newcommand{\DAA}
{\affiliation{David A. Dunlap Department of Astronomy and Astrophysics, 50 St. George Street, University of Toronto, ON M5S 3H4, Canada}}

\newcommand{\STSCI}
{\affiliation{Space Telescope Science Institute, 3700 San Martin Drive, Baltimore, MD 21218, USA}}

\newcommand{\WVUPHAS}
{\affiliation{Department of Physics and Astronomy, West Virginia University, PO Box 6315, Morgantown, WV 26506, USA }}

\newcommand{\WVUGWAC}
{\affiliation{Center for Gravitational Waves and Cosmology, West Virginia University, Chestnut Ridge Research Building, Morgantown, WV 26505, USA}}

\newcommand{\UCB}
{\affiliation{Department of Astronomy, University of California, Berkeley, CA 94720, United States}}

\newcommand{\YORK}
{\affiliation{Department of Physics and Astronomy, York University, 4700 Keele Street, Toronto, ON MJ3 1P3, Canada}}

\newcommand{\PI}
{\affiliation{Perimeter Institute of Theoretical Physics, 31 Caroline Street North, Waterloo, ON N2L 2Y5, Canada}}

\newcommand{\UBC}
{\affiliation{Department of Physics and Astronomy, University of British Columbia, 6224 Agricultural Road, Vancouver, BC V6T 1Z1 Canada}}

\newcommand{\UCHILE}
{\affiliation{Department of Electrical Engineering, Universidad de Chile, Av. Tupper 2007, Santiago 8370451, Chile}}

\author[0009-0007-5296-4046]{Lordrick~A.~Kahinga}
\UCSC
\UDOM

\author[0000-0002-7738-6875]{J.~Xavier~Prochaska}
\UCSC
\IPMU
\NAOJ

\author[0000-0003-3801-1496]{Sunil~Simha}
\altaffiliation{NU-UC Brinson Fellow}
\CIERA
\Uch

\author[0000-0002-4209-7408]{Calvin Leung}
\altaffiliation{NHFP Einstein Fellow}
\UCB

\author[0000-0001-6422-8125]{Amanda~M.~Cook}
\MU
\Trottier
\UVA

\author[0000-0002-1348-8063]{Radu~V.~Craiu}
\DoSS

\author[0000-0003-3734-8177]{Gwendolyn.~Eadie}
\DAA
\DoSS

\author[0000-0001-8384-5049]{Emmanuel Fonseca}
  \WVUPA
  \WVUGWAC

\author[0000-0002-3382-9558]{B.~M.~Gaensler}
\UCSC
\DAA
\DI

\author[0000-0001-9345-0307]{Victoria~M.~Kaspi}
\MU
\Trottier

\author[0009-0004-4176-0062]{Afrokk Khan}
  \MU
  \Trottier

\author[0009-0008-6166-1095]{Bikash Kharel}
  \WVUPA
  \WVUGWAC

\author[0000-0003-2116-3573]{Adam E. Lanman}
    \MITK
    \MITP

\author[0000-0002-7164-9507]{Robert A.~Main}
  \MU
  \Trottier

\author[0000-0003-4584-8841]{Lluis Mas-Ribas}
\UCSC

\author[0000-0002-4279-6946]{Kiyoshi W.~Masui}
  \MITK
  \MITP

\author[0000-0002-8897-1973]{Ayush~Pandhi}
\MU
\Trottier

\author[0009-0008-7264-1778]{Swarali Shivraj Patil}
  \WVUPA
  \WVUGWAC

\author[0000-0002-8912-0732]{Aaron~B.~Pearlman}
\altaffiliation{NASA Hubble Fellow.}
\affiliation{MIT Kavli Institute for Astrophysics and Space Research, Massachusetts Institute of Technology, 77 Massachusetts Ave, Cambridge, MA 02139, USA}
\affiliation{Department of Physics, Massachusetts Institute of Technology, 77 Massachusetts Ave, Cambridge, MA 02139, USA}
\affiliation{Department of Physics, McGill University, 3600 rue University, Montr\'eal, QC H3A 2T8, Canada}
\affiliation{Trottier Space Institute, McGill University, 3550 rue University, Montr\'eal, QC H3A 2A7, Canada}

\author[0000-0002-7374-7119]{Paul Scholz}
  \YORK

\author[0000-0002-6823-2073]{K.~Shin}
\CALTECH

\author[0000-0003-2631-6217]{Seth R.~Siegel}
  \PI
  \MU
  \Trottier

\author[0000-0002-2088-3125]{Kendrick Smith}
  \PI

\author[0009-0001-3334-9482]{Michele~Woodland}
\UCSC

\correspondingauthor{Lordrick A. Kahinga}
\email{lkahinga@ucsc.edu}






\begin{abstract}

The circumgalactic medium (CGM) surrounding galaxies is believed to be a
significant reservoir of baryons, yet its total mass remains poorly
constrained. We present a novel approach to probe the CGM of the Andromeda
galaxy (M31) using fast radio bursts (FRBs) from the CHIME/FRB Catalog 2.
By comparing the dispersion measures (DMs) of 171
FRBs whose sightlines intersect M31's halo (within $r_{\rm vir}= 302\,\mathrm{kpc}$)
to a control sample of 684 FRBs, we estimate the DM
contribution from M31's CGM. We find evidence for an excess DM
of $\delta\mathrm{DM} = 5.9$--$59.6\,\mathrm{pc\,cm^{-3}}$ in the inner halo ($\approx 0$--$151\,\mathrm{kpc}$) and $\delta\mathrm{DM} = 25.7$--$64.6\,\mathrm{pc\,cm^{-3}}$ in the
outer halo ($\approx 151$--$302\,\mathrm{kpc}$). Using a generalized halo model parameterized by a closure
radius $r_{\mathrm{close}}$, we constrain the baryon distribution and infer a best-fit value for $r_{\mathrm{close}}$ to be
$9.2^{+9.9}_{-4.9}\,r_{\rm vir}$ ($1\sigma$), corresponding to a total CGM mass of $M_{b,\mathrm{halo}} = 18.6^{+7.9}_{-8.4} \times 10^{10}\,M_\odot$.
Our results suggest that M31 may harbor a substantial fraction of its cosmic
baryon budget in diffuse, ionized gas. This work demonstrates the potential
of FRBs as a powerful tool for studying the CGM of nearby galaxies, with
future larger samples expected to provide tighter constraints on the baryon
content of galactic halos.

\end{abstract}
\keywords{Circumgalactic medium -- Fast radio bursts -- Andromeda Galaxy -- Galaxy halos -- Dispersion measure}


\section{Introduction}\label{sec:intro}
The circumgalactic medium (CGM) is a diffuse, multiphase gas that surrounds galaxies and extends out to their virial radii and beyond \citep[e.g.,][]{wilde2021}. It carries information regarding the history of a galaxy, both as a reservoir for gas that fuels star formation and as a repository for material ejected by feedback processes \citep[see reviews by][]{Tumlison_CGM_2017, Peroux_Howk_2020_review, hsiao_chen_cgm, FaucherGiguere2023}. The CGM serves as a bridge between galaxies and the more diffuse intergalactic medium, and is therefore a crucial region for the exchange of baryons and metals due to galactic feedback and interactions, playing a significant role in the evolution of galaxies \citep[e.g.,][]{prochaska+probing2011, Lehner_Howk_M31, Tchernyshyov2023, Zhang_2024, medlock_2024}. 

Another key motivation for studying the CGM is 
that it may be the dominant baryonic component of 
galactic halos. Observations of the Cosmic Microwave Background (CMB) \citep[e.g.,][]{Komatsu2011, Planck2020} constrain the universal ratio of baryonic to total matter density, $\Omega_b/\Omega_m \approx 0.158$ \citep{Planck2020}. If galaxies retain their cosmological share of baryons, we expect a halo of total mass \mhalo\ to contain a baryonic mass of $\mbcosmic \equiv \mhalo (\Omega_b/\Omega_m)$, which we refer to as the cosmic baryonic mass \citep[following][]{Chiu_et_al_2018_Baryon_content_in_ICM}.
For Milky Way-mass galaxies (commonly referred to as $L^*$ galaxies, with stellar masses $\sim 10^{10.5}\, M_\odot$ and halo masses $\sim 10^{12}\, M_\odot$), the mass in galaxies comprising stars, dust, and the interstellar medium (ISM) accounts for only a small 
fraction of \mbcosmic, since the total halo mass is dominated by dark matter with baryons comprising at most $\sim16\%$. Yet observations indicate that the stellar and ISM components together typically represent only $\sim20-30\%$ of the expected baryonic mass within the virial radius.
This so-called galactic
missing baryons problem \citep{McGaugh2010,Behroozi_2019} 
implies that either the majority of baryons 
are in the CGM or the processes of galaxy formation
have driven the gas from the system altogether 
\citep[e.g.,][]{illya2024}.
Therefore, assessing the mass of the CGM in
galactic halos lies central to fundamental issues in
galaxy formation.

Despite its well-recognized importance, measuring the
mass of the CGM has proven very challenging.  Part
of the difficulty is due to its 
multiphase nature, i.e., it contains gas at a range
of temperatures, densities and ionization states, namely: cool gas ($T < 10^5\,\mathrm{K}$),  warm-hot gas ($10^5\,\mathrm{K} < T < 10^6\,\mathrm{K}$), and hot phase $T > 10^6\,\mathrm{K}$ \citep[see review by][]{Putman_2012_review}. 
The cool and warm phases are accessible to absorption-line
experiments that utilize
bright background sources (primarily quasars) 
to probe the CGM. 
Measurements by \citet{Tumlinson_2013, Werk_et_al_2013, Bordoloi_et_al_2014, Liang_et_al_2014, werk_et_al_2014, werk_et_al_2016, prochaska_et_al_2017, Chen_CGM_2020, Faerman2020warmhotgas, Faerman_Werk23, Qu_Chen24, Zheng_et_al_2024} have provided valuable constraints on the warm and cool phases of the CGM. However, these measurements are limited by the brightness of the background sources and the number of sightlines available to probe the CGM of interest. Furthermore, mass estimates require significant ionization corrections and/or assumptions on the gas metallicity \citep{werk_et_al_2014,Burchett_et_al_2016}. Nevertheless, current experiments have estimated the cool CGM to contain $\approx30\%$ of \mbcosmic\ in low-redshift, $L^*$ galaxies \citep{werk_et_al_2014}. 

Recent studies tracing O\,\textsc{vi} and Ne\,\textsc{viii} absorption have inferred that the warm-hot phase of the CGM (with temperatures $T \sim 10^5$–$10^6$\,K) may contain an additional $\sim20$–$30\%$ of the baryonic budget \citep{Faerman2020warmhotgas,Faerman_Werk23, Qu_Chen24}. These estimates are sensitive to assumptions about gas ionization conditions and metallicity. Still, they suggest that the warm phase may be at least as massive as the cool component, and more spatially extended.

Measurements of the hot CGM in $L^*$ galaxies have relied
on X-ray observations \citep[e.g.,][]{Anderson_Bregman2010,Bregman_Xray_2018}.
Most recently,
\citet{zhang_xray_cgm2024} utilized observations from the SRG/eROSITA All-Sky Surveys (eRASS:4) to measure the X-ray surface brightness and detect the hot CGM for $L^*$
galaxies. Their best estimates suggest that the hot CGM is less
than $\approx20\%$ of \mbcosmic, 
however these measurements depend on 
assumptions for the metallicity of the CGM, 
as well as the temperature of the halo. 

With measuring the mass of the CGM as our primary
motivation, we have designed an experiment that focuses
on the Andromeda galaxy (M31),
the massive spiral galaxy located at a distance of \distancetoandromeda\ \citep{Riess2012} from Earth. 
The galaxy's rotation curve and satellite dynamics suggest a total halo mass in the range $(1.5$–$3) \times 10^{12}\, M_\odot$ \citep{Van_der_marel_2012, patel_mandel_2023}, 
making it the nearest large galaxy outside the Milky Way. 
M31 has been extensively studied across multiple wavelengths, yielding insights into its stellar disk, bulge, halo, and dark matter distribution \citep{Renda_contrasting_2004, Wada_mapping_2008, Huo_2013}. The stellar disk of M31 has been mapped in extraordinary detail by the Panchromatic Hubble Andromeda Treasury (PHAT) survey \citep{Dalcanton_2012_PHAT}, which resolved over 100 million stars across a significant portion of the galaxy. Subsequent studies in the PHAT series have characterized the spatially resolved structure, stellar populations, and dust content of the galaxy \citep{Williams_PHAT_II_2012, Dalacanton_PHAT_VIII_2015, Lewis_PHAT_XI_2015, Williams_2017}. PHAT XIX \citep{Williams_2017} used color-magnitude diagram fitting of these resolved stars to derive the star formation history of M31's disk. They estimate that a total of $(1.5 \pm 0.2) \times 10^{11}\, M_\odot$ of stars have formed over the galaxy's history, of which $(9 \pm 2) \times 10^{10}\, M_\odot$ have likely survived to the present.
These properties make M31 a natural laboratory for testing whether massive galaxies retain their expected fraction of baryons during their evolution.

Furthermore, 
M31's proximity and therefore large angular extent
in the sky (with a halo subtending over $23^\circ$ corresponding to its virial radius of \virialradius\ )
make it the best candidate to resolve its CGM with 
multiple intersecting sightlines from background sources 
\citep{Lehner_Howk_M31,AMIGA_Lehner_2020}. 
Observational studies using absorption-line spectroscopy of background quasars \citep{Lehner_Howk_M31, howk_amiga_2017, AMIGA_Lehner_2020, AMIGA_2025} indicate that M31's circumgalactic medium (CGM) contains a substantial reservoir of gas. \citet{AMIGA_2025} have probed the cool CGM using Si\,\textsc{ii}, Si\,\textsc{iii}, and Si\,\textsc{iv} absorption lines, and the warm CGM using O\,\textsc{vi}. The detection of O\,\textsc{vi} beyond one virial radius suggests the presence of highly ionized gas extending well into the outer halo. They estimate the baryonic mass of the cool and warm CGM to be $\geq 3.7 \times 10^{10}(Z/0.3Z_\odot)^{-1} \, M_\odot$. Assuming a dark-matter halo mass ($\mhalo=1.54\times10^{12}M_\odot$) \citep{Van_der_marel_2012}; and a metallicity of $0.3Z_\odot$, the mass estimated for warm and cool phases accounts for $\approx30\%$ of the baryons in the CGM of M31.

Regarding the hot gas associated with M31,
\citet{Bregman_Lloyd2007} and more recently \citet{Qu2020HotBridge} combine X-ray emission spectra from highly ionised ions (e.g., O\,\textsc{vi} and O\,\textsc{vii} ) and Sunyaev-Zel'dovich (SZ) and claim a ``baryon bridge'' connecting the M31 and MW galaxies. This structure can be approximated by a cylinder with a radius of $120\,\mathrm{kpc}$ and a length of $400\,\mathrm{kpc}$. They estimate a temperature
of $\log T(\mathrm{K}) = 6.34 \pm 0.03$, a density ranging from $10^{-4}\,\mathrm{cm}^{-3}$ to $10^{-3}\,\mathrm{cm}^{-3}$, a metallicity between $0.02\,Z_\odot$ and $0.1\,Z_\odot$, and a baryon mass of at least $10^{11}\,M_\odot$. However, there are significant uncertainties in the measurements of the SZ signal and the estimated column densities from the X-ray emission spectra of ionised metals.

To advance upon these previous works, we here leverage
a large sample of fast radio bursts (FRBs) derived from 
the CHIME/FRB project \citep{collaboration2018chime}.
FRBs are highly energetic, extragalactic bursts that last only a few milliseconds to microseconds \citep{lorimer2007bright}. 
As they traverse through diffuse and ionized plasma, their light disperses 
with the delay ($\Delta t$) in the FRB pulse 
a function of the pulse frequency $\nu$ 
given by $\Delta t =
4.148808\,\mathrm{ms}
\left( \frac{\Delta\nu}{\mathrm{GHz}} \right)^{-2}
\left( \frac{\mathrm{DM}}{\dmunits} \right)$.
The proportionality constant DM is the 
dispersion measure, which is the sightline 
integration of the electron density weighted by the cosmological scale factor

\begin{equation}
    \mathrm{DM} = \int \frac{n_e}{1+z} \, ds
\end{equation}
 \citep[e.g.,][]{petroff2015, chatterjee2017direct,kulkarni2020dispersion}. 
Of greatest importance to this paper, the DM can be used to infer the electron column density of the diffuse baryons along a sightline through the CGM
independent of their physical condition.

FRBs are a promising tool for probing diffuse matter, especially for baryons in the warm ionized phase, which is barely traceable through self-emission and absorption spectra \citep{mcquinn2014locating,prochaska2019probing,connor_and_Ravi_halo, Wu_McQuinn2023}. While previous studies have demonstrated the utility of FRBs for probing the cosmic distribution of baryons \citep{prochaska2019probing,macquart2020census,2024arXiv240200505K}, these analyses have focused on statistical inferences over large, diverse halo populations. 

In contrast, \citet{Reshma_2025} used two localized FRBs with sightlines passing through the inner M31 halo ($\rperp \lesssim 0.1\,\rvir$) to estimate DM contributions of $\sim 9$--$145\,\dmunits$ and $\sim 28$--$286\,\dmunits$ along individual sightlines, providing constraints on M31's CGM from FRBs.

In this work, we examine the properties of a single, well-characterized system, M31, whose large angular extent and proximity to the Milky Way place it in the path of multiple FRB sightlines. We construct a sample of FRBs that intersect the M31 halo and statistically compare their DMs to a control sample of non-intersecting FRBs. From the inferred DM associated with the M31 halo, we constrain the total ionized gas content of its CGM and, by extension, its baryonic mass.

This paper is structured as follows: In Section~\ref{sec: methodology}, we describe the methods used for modelling the M31 halo, data selection, and construction of a control sample. In Section~\ref{sec: analysis}, we set out the analysis techniques, sandboxing and validating the methods utilising simulated samples of FRBs. In Section~\ref{sec: results}, we apply the developed methodology to the real data and report the results. In Section~\ref{sec: Discussion}, we discuss the implications of our results as they pertain to the total mass of the baryons in the CGM of M31 and the fraction of baryons retained by the halo. 
We compare our methodology with other probes used to study the CGM of M31, and also discuss possible systematics 
and the potential for future work to attain better constraints. Finally, we summarise our conclusions in Section~\ref{sec: Conclusion}.


\section{Methodology}\label{sec: methodology}

\begin{figure*}[htbp]
    \centering
    \includegraphics[width=\textwidth]{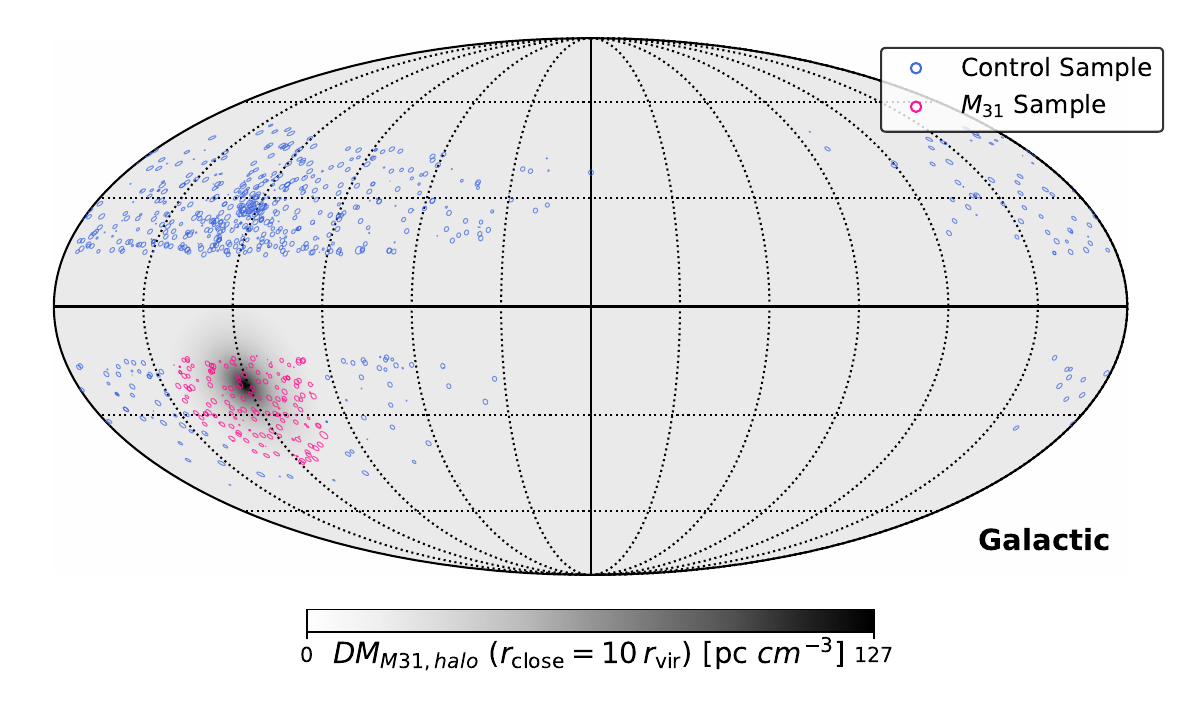}
    \caption{Skymap of the final analysis catalog (built from FRBs in  CHIME/FRB Catalog 2 \citep{collaboration_second_2026}) showing the locations of \numfrbscontrolsampleinhalo\ FRBs intersecting M31's halo (within $r_{\mathrm{vir}}$) and the corresponding control sample of \numfrbscontrolsampleouthalo\ FRBs. 
    The ellipse markers represent the $\approx3\sigma$ localisations of each FRB. The inset in the bottom left region shows the extent of the M31 halo and the predicted \dmtarg\ contribution from a generalized halo model (see Section~\ref{sec:cgm_profile}) with an assumed $\rclose = 10.0~r_{\rm vir}$. The overdensity of detected FRBs at the Celestial North Pole is due to the orientation of CHIME, which is positioned to face North-South and is located in the Northern Hemisphere; therefore, it is always oriented North-South and the North Pole is always in the field of view.}
    \label{fig:control_sample}
\end{figure*}

\subsection{Modeling M31's Halo}\label{sec:modelling_M31}

We adopt the following physical parameters for the M31 system, summarized 
in Table~\ref{tab:m31_params}.

\textbf{Distance and Position:}
We adopt a distance to M31 of \distancetoandromeda\ \citep{Riess2012} and place 
the projected center of the halo at Galactic coordinates 
$(l, b) = (121.17^\circ, -21.57^\circ)$ \citep{Courteau_2011}.

\textbf{Halo Mass:}
We adopt a total halo mass of $\mhalo = 1.5 \times 10^{12}\,M_\odot$, 
as estimated by \citet{Van_der_marel_2012} using Local Group kinematics. 
This is consistent with constraints from M31's rotation curve and 
satellite dynamics \citep{patel_mandel_2023}.

\textbf{Virial Radius:}
We define the virial radius using the spherical collapse overdensity 
for a flat $\Lambda$CDM cosmology \citep{Bryan_Norman_1998}. The virial 
overdensity relative to the critical density is
\begin{equation}
\Delta_{\mathrm{vir}} = 18\pi^2 - 82q - 39q^2,
\label{eqn:delta_vir}
\end{equation}
where $q \equiv \Omega_\Lambda / [\Omega_\Lambda + \Omega_m(1+z)^3]$ is 
the dark energy density parameter at redshift $z$. The virial radius is 
then
\begin{equation}
\rvir = \left( \frac{3\,\mhalo}{4\pi\,\Delta_{\mathrm{vir}}\,\rho_c} \right)^{1/3},
\label{eqn:rvir}
\end{equation}
where $\rho_c = 3H_0^2/(8\pi G)$ is the critical density. For our adopted 
halo mass and $z \approx 0$, this yields $\rvir = \virialradius$, corresponding 
to an angular radius of $\approx 23^\circ$ at the distance of M31.

\textbf{Stellar Mass:}
The stellar mass of M31 is $\mstar = \stellarmass$ 
\citep{Tamm_mass_2012}, corresponding to $\approx 50\%$ of the cosmic 
baryonic mass $\mbcosmic \equiv (\Omega_b/\Omega_m)\,\mhalo \approx 
24 \times 10^{10}\,M_\odot$. This suggests that a substantial fraction 
of baryons may reside in the CGM.

\begin{deluxetable*}{lcl}
\tablecaption{Adopted parameters for the M31 system.\label{tab:m31_params}}
\tablehead{
\colhead{\textbf{Parameter}} & \colhead{\textbf{Value}} & \colhead{\textbf{Reference}}
}
\startdata
Distance $d$            & \distancetoandromeda                 & \citet{Riess2012} \\
Galactic $(l, b)$       & $(121.17^\circ, -21.57^\circ)$       & \citet{Courteau_2011} \\
Halo mass \mhalo        & $1.5 \times 10^{12}\,M_\odot$         & \citet{Van_der_marel_2012} \\
Virial radius \rvir     & \virialradius                         & \citet{Bryan_Norman_1998} \\
Stellar mass \mstar     & $\stellarmass$                        & \citet{Tamm_mass_2012} \\
Cosmic baryon mass \mbcosmic & $2.4 \times 10^{11}\,M_\odot$     & Derived from $\fb$ \\
\enddata
\end{deluxetable*}


\subsubsection{CGM Density Profile}
\label{sec:cgm_profile}

To model the distribution of ionized gas in M31's CGM, we employ a 
generalized halo model that builds on the modified Navarro-Frenk-White (mNFW) 
profile introduced by \citet{prochaska2019probing}. Our framework addresses 
a key limitation of standard mNFW models: they lack a mechanism to enforce 
the cosmic baryon fraction at large radii.

The central insight of our approach is to define the enclosed baryon mass 
as a modulated fraction of the enclosed dark matter mass:
\begin{equation}
M_b(<r) = g(y) \, \frac{\Omega_b}{\Omega_m - \Omega_b} \, M_{\mathrm{dm}}(<r),
\label{eqn:Mb_prescription_main}
\end{equation}
where $y \equiv c\,r/\rvir$ is the dimensionless radius scaled by the 
concentration parameter $c$, and $g(y)$ is a modulation function satisfying 
$g(0) = 0$ (baryon depletion in the inner halo due to feedback) and 
$\lim_{y \to \infty} g(y) = 1$ (recovery of the cosmic baryon fraction at 
large radii). This construction guarantees that the enclosed baryon fraction 
approaches $\Omega_b/\Omega_m$ asymptotically, regardless of the specific 
form of $g(y)$.

We parameterize $g(y)$ using a hyperbolic tangent function:
\begin{equation}
g(y) = \tanh\left(\frac{y}{y_{\mathrm{out}}}\right),
\label{eqn:g_tanh_main}
\end{equation}
where the scale parameter $y_{\mathrm{out}}$ is related to a physically 
motivated \textit{closure radius} $\rclose = k \, \rvir$ at which the enclosed 
baryon fraction reaches the cosmic mean \citep{sorini_how_2022, ayromlou2023}. 
The parameter $k$ thus controls the radial extent of baryon redistribution 
relative to the virial radius.

The baryon density profile derived from Equation~\eqref{eqn:Mb_prescription_main} 
takes the form:
\begin{equation}
\rho_b(y) = \frac{\Omega_b}{\Omega_m - \Omega_b} \rho^0_{\mathrm{dm}} \left[ 
    \frac{g(y)}{y(1+y)^2} 
    + \frac{1}{y^2} \frac{dg}{dy} \left(\ln(1+y) - \frac{y}{1+y}\right)
\right],
\label{eqn:rho_b_main}
\end{equation}
where $\rho^0_{\mathrm{dm}}$ is the NFW normalization density. The first term 
represents a modulated NFW profile, while the second term ensures proper 
asymptotic behavior. The full derivation and discussion of model properties 
will be provided by Simha et al (in prep).

For our fiducial analysis, following \citet{prochaska2019probing}, we adopt $c = 7.67$ for the NFW concentration 
parameter. We treat $k$ as a free parameter to be 
constrained by the FRB observations, exploring values in the range 
$k = \kboundlower$--$\kboundupper$ corresponding to closure radii from $\kboundlower\,\rvir$ to 
$\kboundupper\,\rvir$. Table~\ref{tab:halo_model_params} summarizes the model parameters.

\begin{table}[htbp]
\centering
\caption{Parameters of the generalized halo model.}
\label{tab:halo_model_params}
\begin{tabular}{lcl}
\toprule
\textbf{Parameter} & \textbf{Value/Range} & \textbf{Description} \\
\midrule
$c$ & 7.67 & NFW concentration parameter \\
$k$ & \kboundlower--\kboundupper & Closure radius in units of $\rvir$ \\
$\Omega_b/\Omega_m$ & 0.158 & Cosmic baryon fraction \\
\bottomrule
\end{tabular}
\end{table}


\subsubsection{Dispersion Measure from M31's Halo}

The dispersion measure contribution from M31's CGM at impact parameter 
\rperp\ is obtained by integrating the free electron density along the 
line of sight:
\begin{equation}
\dmtarg(\rperp) = 2 \int_0^{\sqrt{\rclose^2 - \rperp^2}} n_e(\ell)\, d\ell,
\label{eqn:dm_m31}
\end{equation}
where the factor of 2 accounts for both the near and far sides of the halo, 
and the integration extends to the closure radius $\rclose = k\,\rvir$.

The electron density is related to the baryonic mass density by:
\begin{equation}
n_e = \frac{\mu_e\,\rho_b}{m_p\,\mu_H},
\label{eqn:n_e_main}
\end{equation}
where $m_p$ is the proton mass, $\mu_H = 1.3$ is the mean mass per 
hydrogen atom (accounting for helium), and $\mu_e = 1.167$ is the 
number of free electrons per baryon assuming full ionization with 
primordial helium abundance.

These predicted DM contributions are comparable to the scatter in the 
FRB DM distribution from other sources (e.g., \dmcosmic\ and \dmhost; see Section \ref{sec:dispersion_measures}), 
motivating the statistical approach developed in Section~\ref{sec: analysis}.

\subsection{FRB Selection}\label{sec:data_selection}
Our analysis utilizes the CHIME/FRB Catalog 2 \citep{collaboration_second_2026}, which includes 3641 unique FRB sources detected by the CHIME experiment, as described by \citet{collaboration2018chime}. Briefly, CHIME is a cylindrical radio transit telescope located near Penticton, British Columbia. The telescope comprises four semicylindrical paraboloid reflectors, each measuring 20 meters in diameter and 100 meters in length, and is equipped with 256 dual-polarization feeds. Using a robust backend system, the CHIME/FRB project processes real-time data to search for FRBs within its field of view. CHIME detects $2$--$3$ FRBs per day on average \citep{josephy2021no}, an exceptionally high detection rate made possible by its large field of view, wide bandwidth, high sensitivity, and powerful correlator.

\citet{Cook24} model the sensitivity of CHIME/FRB as a function of source declination ($\delta$), parameterized by the relation $\cos^\iota(\phi - \delta)$, where $\phi$ is the telescope's geographic latitude and $\iota$ is empirically fit using CHIME/FRBs beam model as a prior\footnote{Available at \url{https://chime-frb-open-data.github.io/beam-model/}}. They find \(\iota \approx 1.43\), indicating a steep decline in sensitivity with increasing zenith angle. The peak sensitivity occurs near $\delta \approx 49.3^\circ$, placing M31 (at $\delta \approx 41.3^\circ$) close to this maximum.

An event is deemed extragalactic if its \dmt\ exceeds the maximum Galactic estimate for that sightline (\dmism) predicted by models \citep[][]{cordes2002ne2001,Yao_2017} by at least $5\sigma$. 
Catalog 2 encompasses all bursts detected between 25 July 2018 and 15 September 2023, including 536 bursts from CHIME/FRB Catalog 1 \citep{collaboration2021chime}. In our analysis, we include all non-repeating bursts and the first burst from each apparent repeater in Catalog 2; using the first burst from repeaters avoids double-counting baryons along the same sightline, while selecting the first burst ensures we use detections with the highest signal-to-noise ratio threshold \citep{fonseca2020nine,pleunis2021fast}. We have excluded 13 FRBs 
detected during commissioning and deemed unreliable for statistical analysis. 
For the sky positions of the FRBs, we used the header localization positions found in the catalog. These positions come with an uncertainty of approximately $\approx0.2\times0.2\deg^2$ at a $1\sigma$ confidence. Compared to the M31 halo, which spans $30\deg$, we can safely ignore the uncertainties in the FRB positions 
(see Figure~\ref{fig:control_sample}).
The catalog contains \numfrbsfullsample\ sightlines, with \numfrbsfullsampleinhalo\ 
traversing the M31 halo defined by \rvir. 

\subsection{Dispersion Measures}\label{sec:dispersion_measures}

For each FRB, we adopt the dispersion measure
obtained by the {\sc fitburst} algorithm \citep[for details, see][]{collaboration2021chime}, \dmt\.
The observed \dmt includes all components along the sightline, as described by Equation \ref{eqn:dm_def}. 

\begin{equation}\label{eqn:dm_def}
    \dmt = \dmism + \dmmwhalo + \dmcosmic + \dmhost
\end{equation}
which is the combined  contribution from the Milky Way's interstellar medium (\dmism), the contribution from the Milky Way halo (\dmmwhalo), and the contributions from the
cosmic web (\dmcosmic) and the FRB host galaxy \dmhost.
Both of these latter components are functions of the FRB redshift, \zfrb. 
In this paper, we define the Milky Way-ISM subtracted DM:

\begin{equation}\label{eqn:dm_eg_def}
    \dmeg \equiv \dmt - \dmism
\end{equation}
with \dmism\ values given by 
the NE2001 model \citep{cordes2002ne2001}.
For sightlines that intersect the M31 halo, \dmeg\ will include the
contribution from \dmtarg\ whereas we expect a negligible contribution
at separations greater than \rvir.
This enables us to perform a differential experiment between a sample
of M31 FRBs and a matched control sample.

To minimize uncertainties arising from \dmism, we excluded all FRBs located near the Galactic plane, specifically those within  $\pm5^\circ$ of the Galactic plane ($b = 0^\circ$).
Additionally, we removed all sightlines with $\dmism > 100\,\dmunits$. 
Throughout this paper, we shall refer to FRBs whose sightlines intersect the M31 halo within the virial radius, which is $\approx 23^\circ$ from the center of M31, as M31 FRBs. 
There are \numfrbscleansampleinhalo\ FRBs satisfying the above cuts; these
comprise the final sample of M31 FRBs.

X-ray emission and absorption measurements firmly establish the presence of a hot, extended Milky Way halo \citep{Bregman_Lloyd2007}. However, its density structure and total column remain uncertain. \citet{keating2020exploring} compared a suite of gas-density profiles with soft X-ray constraints and showed that the resulting \dmmwhalo\ spans more than an order of magnitude across physically allowed models. Because of this large model-dependent uncertainty, and the additional anisotropy introduced by our off-center position in the Galaxy, we do not attempt to model \dmmwhalo\ explicitly. Instead, our experiment is differential: by comparing sightlines that intersect the M31 halo with a declination- and latitude-matched control sample, we effectively remove the mean Milky Way halo contribution and isolate the \dmtarg\ associated with M31.

\begin{figure}[htbp]
    \centering
    \includegraphics[width=\columnwidth]{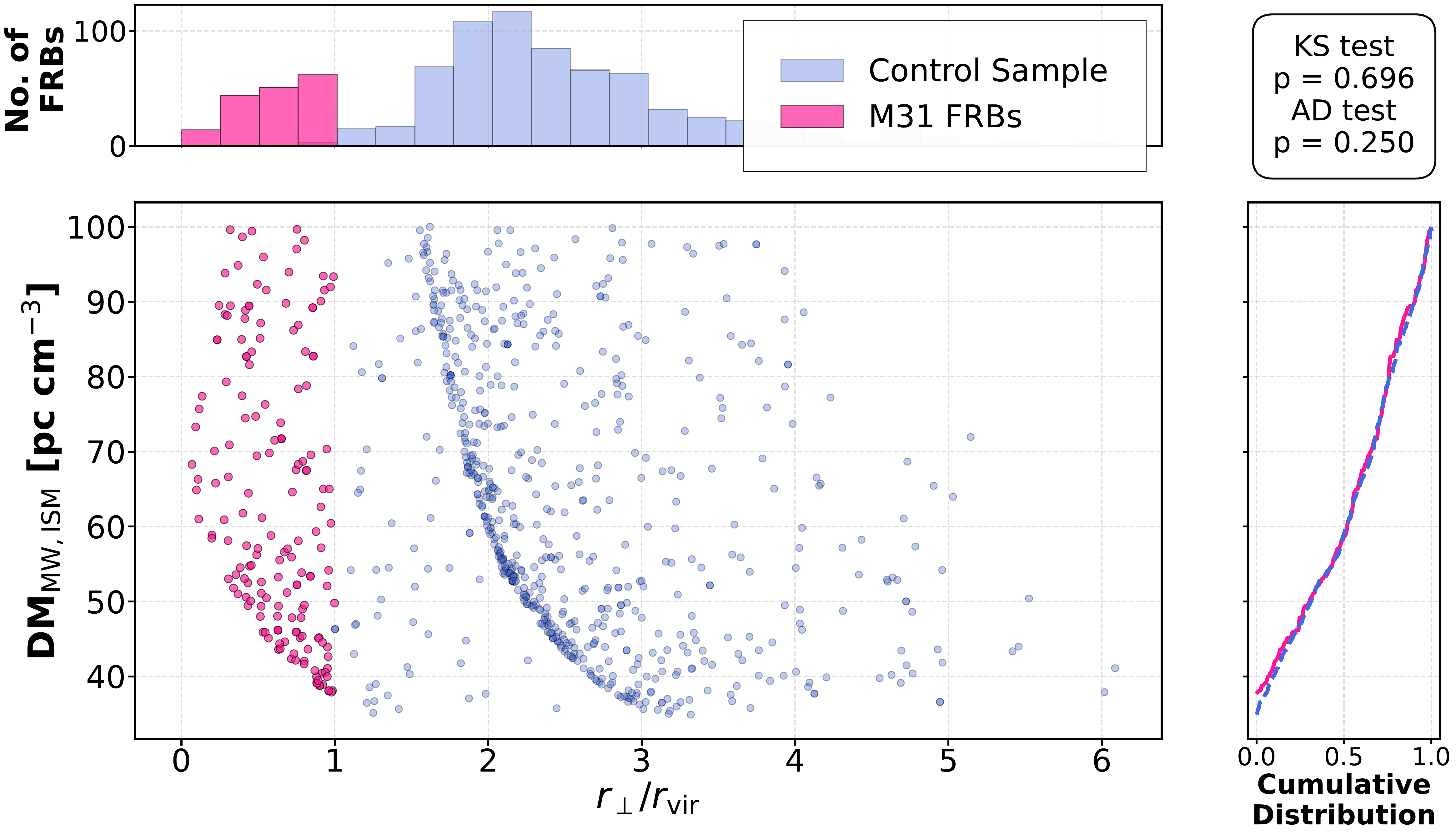}
    \caption{Top: Number of FRBs in the M31 (pink) and control (blue) samples
    as a function of sightline impact-parameter from M31 (\rperp). 
    By design, the number of FRBs in the control sample is four times that of M31 sightlines. Lower left: A comparison of the \dmism\ values in \rperp\ space between the 
    M31 sightlines and those for the control sample.  
    Lower right: A CDF plot comparing the \dmism\ distributions of the control sample vs the M31 sightlines, indicating they were drawn from the same distribution. We performed an Anderson-Darling (AD) test to compare the control sample vs the M31 FRBs sample in \dmism. The AD test returned a $p$-value of \adtestdmismpvalue, indicating we cannot reject the null hypothesis that the two are from the same distribution.} 
    \label{fig:control_sample_hist}
\end{figure}

\subsection{Control Sample} \label{sec:control_sample}
An important challenge in isolating \dmtarg\ is the need to properly account for uncertainty incurred by line-of-sight variations in \dmism. Since the ISM structure is not uniform across the sky, particularly at low Galactic latitudes, spatially correlated fluctuations in \dmism\ can bias inferences drawn from comparing FRBs intersecting M31 with those that do not. This is especially problematic when using a global or averaged model of the Milky Way contribution. To address this, we construct a control sample of non-M31 FRBs that are matched in key sky properties to the M31 sightlines, specifically in \dmism\ and Galactic latitude, thereby minimizing differential systematics. 
This strategy allows us to attribute observed DM differences more 
confidently to the CGM of M31.

The control sample is built from the set of FRBs beyond M31's virial radius
as follows. 
For each FRB in the M31 sample, we
select corresponding FRBs from the sightlines away from the 
M31 halo that satisfies the following criteria:

\begin{itemize}
    \item The \dmism\ values must match within 
    $\pm3 \, \dmunits$. 
    This minimizes the statistical difference in the \dmism\ distribution. 
    \item The Galactic latitudes must match within 
    $|\delta b| < 5$\,deg. This is intended to further
    reduce differences in the Galactic ISM contributions to the two samples.
    Since the \dmism\ distribution in Galactic latitude is expected to be symmetric except for individual HII regions, FRBs in a similar Galactic latitude in the Northern hemisphere also satisfy this condition.  Indeed, Figure~\ref{fig:control_sample}
    shows that the control sample is biased towards the Northern hemisphere,
    a potential systematic we discuss in Section~\ref{sec: systematics}.
\end{itemize}

We repeat the above procedure for all M31 FRBs while ensuring each M31 FRB has the same number of FRBs from non-M31 sightlines associated with it so as not to alter the entire sample's PDF.  With the above criteria, each M31 FRB could be matched to at least 4
non-M31 FRBs;  these \numfrbscontrolsample\ FRBs comprise 
the control sample. 

To validate the similarity between the \(\mathrm{DM_{ISM}}\) distributions of the final analysis samples, we performed a two-sample Anderson-Darling (AD) test. The AD test yielded a $p$-value of \adtestdmismpvalue, indicating that the M31 and non-M31 \(\mathrm{DM_{ISM}}\) samples are statistically consistent with being drawn from the same underlying distribution. 
Figure \ref{fig:control_sample_hist} shows the resulting CDF plot of the control sample vs. M31 FRBs and illustrates each sample's sizes; the control sample size is four times larger than the M31 FRBs sample, 
ensuring significant statistical power for assessing the average sightline away
from M31. 
Figure~\ref{fig:control_sample_hist} also shows the distribution of the two samples as a function of distance from the center of M31, reported as the fraction of the impact parameter to the virial radius of M31 (see Section \ref{sec:modelling_M31}). 

The distribution of our final analysis catalog in Galactic coordinates is illustrated in Figure~\ref{fig:control_sample}. The final sample consists of 
\numfrbscontrolsampleinhalo\ M31 FRBs, along with a control sample of \numfrbscontrolsampleouthalo\ FRBs. Among the M31 sightlines, \numfrbscontrolsampleinhaloinnerbin\ lie within half the virial radius, while \numfrbscontrolsampleinhalouterbin\ lie in the outer half.
Only four sightlines are within 30 kpc of M31's center, such that our estimates exclude diffuse and ionized baryons from the M31 disk.


\FloatBarrier
\section{Analysis Framework} \label{sec: analysis}

\subsection{Sandboxing with Mock FRB Samples}
\label{sec: mock sample generation}

To develop and test our analysis techniques,
we construct a ``sandbox'' of mock FRBs with a known
M31 halo contribution to their DMs.
The two main ingredients in the mock samples are the sky positions and the \dmt\ of the FRBs.

To generate mock sky positions consistent with the spatial distribution of FRBs in CHIME/FRB Catalog~2 \citep{collaboration_second_2026}, we fit a two-dimensional kernel density estimate (KDE) to the joint distribution of right ascension (RA) and declination (Dec) from the full catalog (prior to any selection cuts). We then draw mock (RA, Dec) pairs from this KDE, yielding sky positions that faithfully reproduce the correlated spatial structure of the observed FRB population, including CHIME's declination-dependent sensitivity.

Similar to real data, we assume the \dmism\ values for the mock sightlines
follow the NE2001 model and assign them according to their position on the sky. To speed up the calculation of \dmism, we constructed a HEALPix map (resolution $\approx 0.01\deg$) of its values
across the sky.
For each mock FRB, the corresponding \dmism\ is
assigned by querying this map at the generated sky position.

We generate mock values of \dmt\ using inverse transform sampling. We construct a cumulative distribution function (CDF) of the \dmt\ distribution from the real data, draw uniform random variates from the interval $[0, 1]$, and interpolate through the CDF to obtain mock \dmt\ values. The extragalactic contribution is then computed as $\dmeg = \dmt - \dmism$.
We then proceed to perform \dmism\ and Galactic latitude cuts
(see Sections~\ref{sec:data_selection}
and \ref{sec:control_sample})
to make the final mock sample for analysis. 

From the surviving set of FRBs that intersect M31, 
we generate a control sample of 4~FRBs per M31 FRB.
At this stage, the mock M31 sightlines and control sample are drawn from the same underlying \dmeg\ distribution, but random sampling fluctuations can introduce a spurious offset between their weighted means. To ensure an unbiased baseline before injecting a mock signal, we randomly replace the \dmeg\ values for the M31 sightlines with values drawn from the control sample until 
\ddM\ (the difference in the weighted mean \dmeg\ between the M31 and control samples; defined formally in Section~\ref{sec:weighted mean diff}) 
is less than $10^{-3}\,\dmunits$ (i.e., negligible compared to the expected halo signal of $\mathcal{O}(50\text{--}200)\,\dmunits$\citep[e.g.,][]{prochaska2019probing, connor_and_Ravi_halo, Wu_McQuinn2023}), where \ddM\ serves as our primary estimator of the M31 halo's DM contribution. This procedure guarantees that any non-zero \ddM\ measured after signal injection is attributable solely to the injected M31 contribution, rather than to pre-existing statistical fluctuations between the samples.
To the M31 FRB sightlines, we then add the \dmtarg\ signal we hope to extract; we assume an M31 CGM described by the generalized halo model defined in Section~\ref{sec:cgm_profile}.

We created mock samples of varying sizes to test our methodology. Our primary mock sample contains 3,000 FRBs (henceforth \mocksampleog), including 246 M31 sightlines, mirroring the sample size of the CHIME/FRB Catalog~2.
After applying the data selection criteria, this yields \nummockfrbsinhalosmall~mock M31 FRBs.
To assess how our methodology would perform with larger samples and to evaluate its predictive power, we also created additional samples with $\approx$20,000 and 500,000 FRBs
(see Section~\ref{sec:future} for further discussion).

\subsection{Unweighted Comparison}
\label{sec:ks_analysis}

As a first test on whether the \dmeg\ values of the mock M31 FRBs
differ statistically from the mock control sample, we may compare their
cumulative distribution functions (CDFs), shown in 
Figure~\ref{fig:mock_ks_test}. 
We obtain these CDFs from the \mocksampleog\ sample, after applying Galactic plane and \dmism\ cuts. The sample contained 186 M31 FRBs (mirroring the 177 M31 FRBs in the real sample). The blue curve illustrates the CDF of the 
\dmeg\ of the control sample, while the magenta curve represents the CDF of the \dmeg\ of M31 FRBs, which includes a mock signal added
as described in Section~\ref{sec: mock sample generation}.
To assess whether the apparent excess in \dmeg\ values for the mock M31 FRBs
is statistically significant, we perform a two-sample Anderson-Darling (AD) test on the two CDFs. 
The null hypothesis assumes that both samples are drawn from the same underlying distribution. 
Because our M31 and control samples are imbalanced in size, we compute $p$-values using a permutation test rather than relying on asymptotic distributions. We pool the two samples, randomly partition them into groups matching the original sample sizes, and recompute the test statistic for each of $B = 1{,}000$ permutations. The $p$-value is then given by:

\begin{equation*} \label{eq:permtest}
    p = \frac{1 + \sum_{b=1}^{B} \mathbf{I}[T_b \geq T_{\mathrm{obs}}]}{(B + 1)}    
\end{equation*}

Here $T_{\mathrm{obs}}$ is the AD test statistic computed from the observed samples, $T_b$ is the statistic computed from the $b$-th random partition of the pooled data, and $\mathbf{1}[T_b \geq T_{\mathrm{obs}}]$ is the indicator function, equal to 1 when the permuted statistic equals or exceeds the observed value and 0 otherwise.

This procedure is used for all AD $p$-values reported throughout this paper.
The AD test yields $p$ = \adtestdmegmockpvalue. The test does not detect the injected mock signal at the $p < 0.05$ significance level, illustrating the challenge of detecting M31's DM contribution using a simple CDF comparison of the full sample.

The downside to this simple statistical comparison is that one cannot estimate any quantitative difference
in \dmeg\ between the two samples.  
Furthermore, the high-DM events likely include substantial contributions 
unrelated to M31, reducing sensitivity to a halo signal. 
This motivates the use of a \textit{weighted mean} approach, 
as introduced in the following section.

\subsection{Weighted Estimator of the M31 Contribution (\texorpdfstring{\ddM}{deltaDM})} 
\label{sec:weighted mean diff}

\citet{Wu_McQuinn2023} estimated the DM contribution of the CGM 
from low-redshift galaxies with masses 
of $10^{11}$--$10^{13}\,M_\odot$ using CHIME/FRB Catalog 1. 
The sample contained 20–30 FRBs intersecting halos at $< 1\,r_{\mathrm{vir}}$, with most FRBs interacting with only one halo. 
Their analysis introduced a weighted ``stacking scheme''  
to reduce variance and mitigate bias from high-DM FRBs. 
They then estimated the DM contribution from the CGM of 
galaxies as the difference in the weighted mean \ddM\ between FRBs with 
intersecting sightlines and those without. 

We adopt the formalism of \citet{Wu_McQuinn2023} for the CGM of M31, 
defining the average M31 contribution as the difference in the 
weighted mean between the \dmeg\ of the M31 sightlines and 
the \dmeg\ of the control sample:

\begin{equation} \label{eq:mean_diff}
    \ddM = \sum_{i \in M_{31}} DM_i \, \hat{w}\left(DM_i\right) - \sum_{i \in cs} DM_i \, \hat{w}\left(DM_i\right)
\end{equation}
where the sums run over FRBs in the M31 sample and control sample (CS), respectively. The normalized weights are defined as
\begin{equation}
\hat{w}(DM_i) = \frac{w(DM_i)}{\sum_j w(DM_j)},
\label{eqn:normalized_weight}
\end{equation}
where the sum in the denominator runs over all FRBs in the respective sample (M31 or CS), and
\begin{equation}
w(\mathrm{DM}) = e^{-(\mathrm{DM}/\alpha)^\beta}
\label{eqn:weight}
\end{equation}
is the unnormalized weighting function with parameters $\alpha$ (a characteristic DM scale) and $\beta$ (controlling the steepness of the weighting). The normalization ensures that the weights within each sample sum to unity, so that Equation~\ref{eq:mean_diff} represents the difference between two weighted mean DM values. We use a similar weighting scheme to \citet{Wu_McQuinn2023}, downweighting FRBs with high \dmeg\ values, which are more likely to contain large contributions from the intergalactic medium and other intervening halos.

The weighting function assigns greater weight to low-DM sightlines, where the M31 contribution constitutes a larger fraction of the total \dmeg. The parameter $\alpha$ sets the DM scale at which the weight drops to $1/e$, effectively defining a soft cutoff. The parameter $\beta$ controls how sharply the weights decrease: $\beta = 0$ corresponds to uniform weighting (unweighted mean), while larger $\beta$ produces steeper cutoffs.

Critically, Equation~\ref{eq:mean_diff} is an unbiased estimator only when $\beta = 1$ \citep{Wu_McQuinn2023}. For $\beta \neq 1$, the estimator introduces a systematic offset in \ddM. We therefore fix $\beta = 1$ throughout our analysis to ensure no systematic offset in our estimates.

The optimal value for $\alpha$ is then a trade-off between choosing
lower values,
which minimize the variance in \ddM, and larger values, which include signal
from a larger fraction of the sample.
Plausible DM contributions
from $10^{12}\,M_\odot$ galaxy halos range over
$50$--$200\,\dmunits$ \citep[e.g.,][]{prochaska2019probing,connor_and_Ravi_halo,Wu_McQuinn2023}.
Therefore, we choose to consider values of $\alpha \geq 200\,\dmunits$.

To determine the optimal $\alpha$, we calculated the standard deviation of the distribution of \ddM\ values (\ddmstandardev) for randomly selected FRBs at fixed $\beta = 1$. From our largest mock sample of 500,000 FRBs (see Section~\ref{sec: mock sample generation}), we randomly selected 200 FRBs, similar to the number of M31 sightlines in our real sample, and then calculated \ddM. We repeated this process to create a distribution of 10,000 \ddM\ values for each $\alpha$ and calculated \ddmstandardev. Figure~\ref{fig:alpha_variance} shows how \ddmstandardev\ varies with $\alpha$ at $\beta = 1$. The variance increases monotonically with $\alpha$, rising steeply at low $\alpha$ before flattening at higher values. Small $\alpha$ values ($\lesssim 200\,\dmunits$) yield the lowest variance but aggressively downweight most of the sample, potentially discarding genuine M31 signal. We adopt a fiducial value of $\alpha = 300\,\dmunits$, which achieves a \ddmstandardev\ of $\sim 15\,\dmunits$, representing a reduction by a factor of $\sim 1.6$ compared to the unweighted limit ($\alpha \to \infty$), while retaining sensitivity to the expected M31 halo contribution. The shaded band in Figure~\ref{fig:alpha_variance} shows that our results are insensitive to small variations in $\beta$ around unity ($\beta \in [0.8, 1.2]$). We have verified that our main results are not sensitive to this choice of $\alpha$: varying $\alpha$ within the range $200$--$500\,\dmunits$ does not significantly change our conclusions. We present an exploration of different $\alpha$ values in Appendix~\ref{app:alpha_sensitivity}.


\subsection{Mock Estimates of \texorpdfstring{\ddM}{deltaDM}}
\label{sec:param_modeling_sandbox}

In this section, we validate our analysis framework using the mock FRB samples 
described in Section~\ref{sec: mock sample generation}. We demonstrate that both 
non-parametric and parametric approaches can successfully recover an injected 
M31 halo signal. For these tests, we use the \mocksamplebig\ sample containing 
approximately 1,000 M31 sightlines after quality cuts, and inject a signal 
corresponding to the generalized halo model (Section~\ref{sec:cgm_profile}) 
with a fiducial closure radius of $k = 5.0$ (i.e., $\rclose = 5.0\,\rvir$). 
This choice produces DM contributions in the range $\sim 20$--$150\,\dmunits$ 
across the halo, consistent with predictions from previous CGM studies 
\citep{prochaska2019probing, connor_and_Ravi_halo, Wu_McQuinn2023}.

\subsubsection{Non-parametric, Binned \texorpdfstring{\ddM}{deltaDM} Estimation}
\label{sec:nonparam_mock}

We first develop a non-parametric approach to estimate the radial DM 
contribution of M31, denoted $\ddM(\rperp)$, without assuming a specific 
functional form for the halo profile. Our method calculates \ddM\ in discrete 
bins of impact parameter \rperp, constructing matched M31 and control samples 
for each bin independently. We then compute \ddM\ for each bin using 
Equation~\ref{eq:mean_diff} with our fiducial weighting parameters 
$(\alpha, \beta) = (300\,\dmunits, 1)$.

Figure~\ref{fig:mock_nonparam} shows the recovered \ddM\ profile from the 
\mocksamplebig\ sample divided into five radial bins spanning $0$--$1\,\rvir$. 
The data points represent the measured \ddM\ in each bin, with horizontal 
error bars indicating the bin width and vertical error bars representing 
the $1\sigma$ uncertainty. The magenta shaded region shows the expected 
scatter from random sky positions (i.e., the null hypothesis of no M31 signal), 
estimated by repeatedly drawing matched samples from the control population.

Uncertainties in \ddM\ are estimated using a Monte Carlo approach. For each 
radial bin, we randomly draw the same number of FRBs from the control sample 
as in the corresponding M31 bin and compute an ``off-halo'' \ddM\ value using 
the same estimation procedure. This process is repeated 10,000 times to generate 
a distribution of off-halo \ddM\ values. As expected, these distributions are 
approximately Gaussian with a mean near $0\,\dmunits$, since the control sample 
contains no M31 DM contribution by construction. The standard deviation of each 
distribution is adopted as the $1\sigma$ uncertainty for the corresponding 
\ddM\ measurement.

All five radial bins show significant detections of the injected signal, 
with the measurements marked by ``$***$'' indicating $>3\sigma$ significance. 
The recovered profile decreases monotonically with increasing impact parameter, 
as expected for a centrally concentrated halo. The innermost bin 
($\rperp < 0.2\,\rvir$) recovers $\ddM \approx 150\,\dmunits$, while the 
outermost bin ($0.8 < \rperp/\rvir < 1.0$) yields $\ddM \approx 20\,\dmunits$, 
both consistent with the input model predictions (i.e., following the \ddM corresponding to a generalised halo profile with \rclose $=10\rvir$).

This non-parametric method has several advantages: it requires no assumptions 
about the functional form of the M31 halo profile, and it is flexible with 
respect to the choice of binning scheme. However, the choice of bin boundaries 
is arbitrary, and the method requires a sufficiently large number of FRBs per 
bin to suppress statistical noise. With smaller samples, the inferred \ddM\ 
values become highly uncertain.

\subsubsection{Parametric Modeling of the M31 Halo}
\label{sec:param_mock}

We now develop a parametric approach to constrain the closure radius \rclose\ 
of the generalized halo model directly from the FRB data. This method uses 
all M31 sightlines simultaneously, avoiding the information loss inherent 
in binning, and provides quantitative constraints on the model parameters.

Following the methodology introduced by \citet{Wu_McQuinn2023}, we define 
a goodness-of-fit metric:
\begin{equation}
\xi^2 = \left( \left\{ \sum_{i \in M_{31}}\left( \dmi - \dmg \right) w \left( \dmi - \dmg \right)\right\} - \dmcs \right)^2
\label{eqn:xi2}
\end{equation}
where $\mathrm{DM}_\gamma(\rperp)$ is the model-predicted \dmtarg\ at 
impact parameter \rperp\ for a given value of \rclose, $\mathrm{DM}_i$ is the 
observed \dmeg\ for each M31 sightline, $\overline{\mathrm{DM}}_{\mathrm{cs}}$ 
is the weighted mean DM of the control sample, and $w$ is the weighting 
function defined in Equation~\ref{eqn:weight}.

To evaluate the model fit, we normalize $\xi^2$ by its expected value
$\langle \xi^2 \rangle$ to obtain $\chi^2 = \xi^2 / \langle \xi^2 \rangle$.
The quantity $\langle \xi^2 \rangle$ accounts for the statistical variance
expected from sampling fluctuations and is computed empirically: we draw
$N = 10{,}000$ random subsets of size $n_{\mathrm{FRB}}$ (the number of M31
sightlines) from the control sample, compute the weighted mean DM for each
subset, and calculate the variance of these weighted means relative to the
full control sample mean.

\begin{figure*}[htbp]
    \centering
    \includegraphics[width=0.9\textwidth]{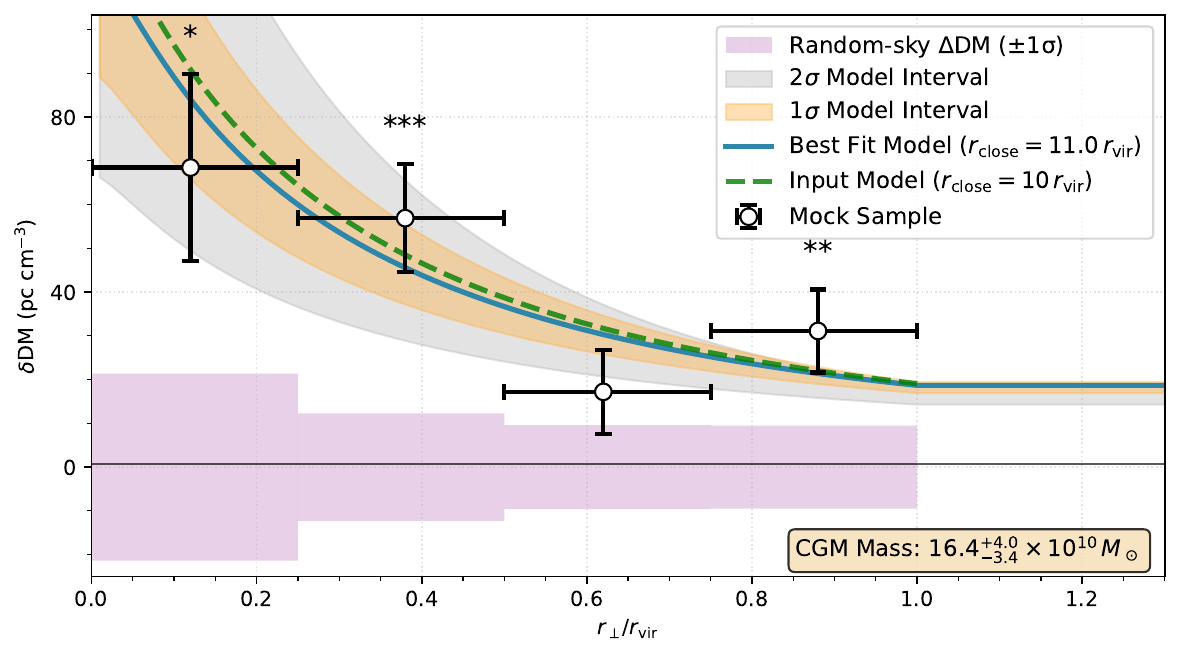}
    \vspace{0.3cm}
    \includegraphics[width=0.9\textwidth]{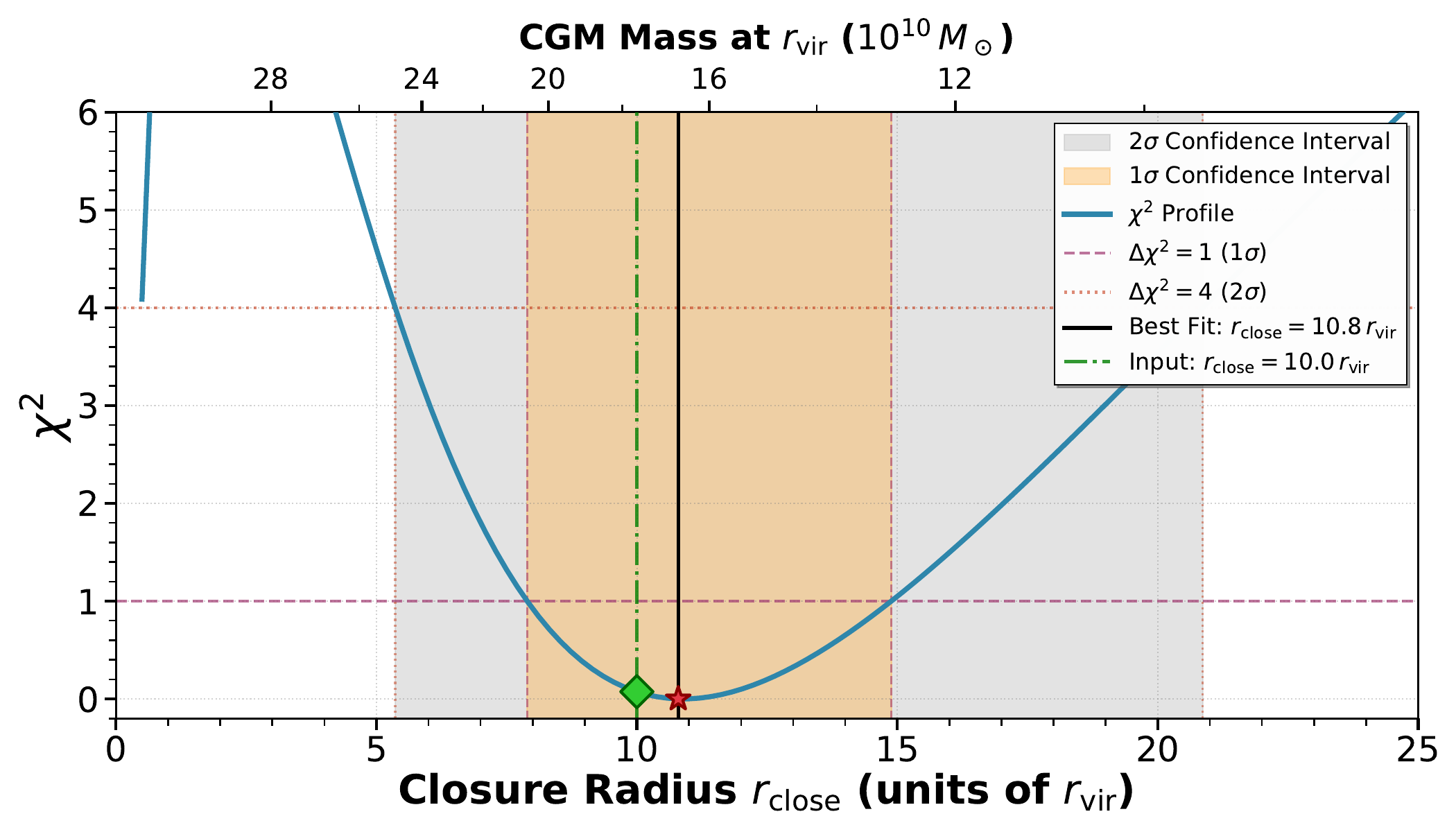}
    \caption{\textbf{Top:} Non-parametric recovery of the injected M31 halo signal from
    the \mocksamplebig\ sample with overplotted best fit values from the parametric analysis. Data points show the measured \ddM\ in five radial
    bins, with vertical error bars indicating $1\sigma$ uncertainties estimated
    via Monte Carlo resampling of the control sample. Horizontal bars indicate
    bin widths. The magenta shaded region shows the $\pm 1\sigma$ range expected
    for random sky positions with no M31 signal. All bins show highly significant
    detections ($***$ indicates $> 3\sigma$). The solid blue curve shows the
    best-fit parametric model ($r_{\mathrm{close}} \approx 11\,\rvir$), with orange
    ($1\sigma$) and gray ($2\sigma$) shaded regions indicating model uncertainties.
    The dashed green curve shows the input model ($r_{\mathrm{close}} = 10\,\rvir$).
    \textbf{Bottom:} Parametric constraints on the closure radius \rclose\
    from the \mocksamplebig\ sample. The blue curve shows the $\chi^2$ profile as a
    function of \rclose\ (in units of \rvir). The horizontal dashed
    lines mark $\Delta\chi^2 = 1$ and $\Delta\chi^2 = 4$, corresponding to $1\sigma$
    and $2\sigma$ confidence levels. The orange and gray shaded regions indicate
    the corresponding confidence intervals. The green diamond marks the input model
    value ($k = 10$), while the red star marks the best-fit value ($k \approx 11$).
    The close agreement validates our parametric framework.}
    \label{fig:mock_nonparam}
    \label{fig:mock_chisq}
\end{figure*}

Figure~\ref{fig:mock_chisq} shows the $\chi^2$ profile as a function of 
\rclose\ for the \mocksamplebig\ sample. The input model used to 
generate the mock signal is marked by the green diamond at $k = 5.0$. The 
best-fit value, corresponding to the minimum $\chi^2$, is $\rclose = 5.5\,\rvir$ 
(red star), in excellent agreement with the input. The $1\sigma$ and $2\sigma$ 
confidence intervals are estimated from $\Delta\chi^2 = 1$ and $\Delta\chi^2 = 4$, 
respectively, yielding approximate ranges of $k \approx 3$--$8$ ($1\sigma$) and 
$k \approx 1.5$--$12$ ($2\sigma$).

The parametric best-fit model is overlaid on the non-parametric measurements 
in Figure~\ref{fig:mock_nonparam}. The solid blue curve shows the best-fit 
model ($k = 5.5\,\rvir$), while the dashed green curve shows the input model 
($k = 5.0$). The orange and gray shaded regions indicate the $1\sigma$ and 
$2\sigma$ model uncertainties, respectively. The agreement between the 
non-parametric binned estimates and the parametric model fit demonstrates 
the internal consistency of our analysis framework and validates our ability 
to recover injected halo signals from FRB samples of this size.

Once \rclose\ is constrained, the total CGM mass can be derived under the 
assumptions of the generalized halo model. For a given \rclose, the enclosed 
baryon mass within \rvir\ is determined by the model prescription 
(Equation~\ref{eqn:Mb_prescription_main}). We defer the application of 
this framework to the real data to Section~\ref{sec: results}.


\begin{figure*}[htbp]
    \centering
    \includegraphics[width=\textwidth]{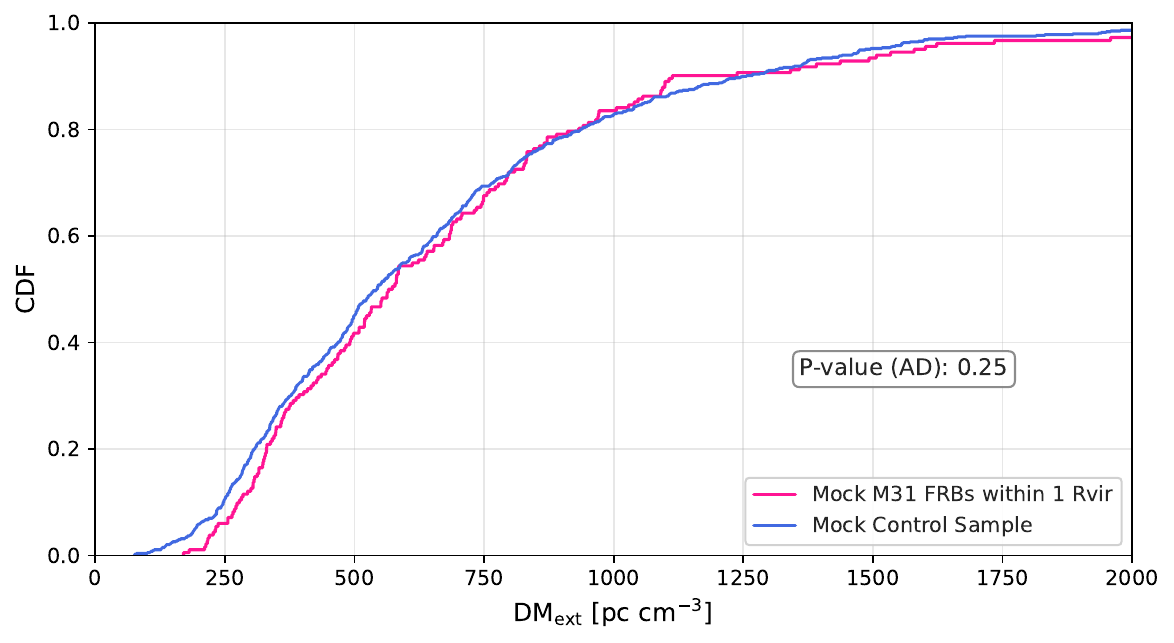}
    \includegraphics[width=\textwidth]{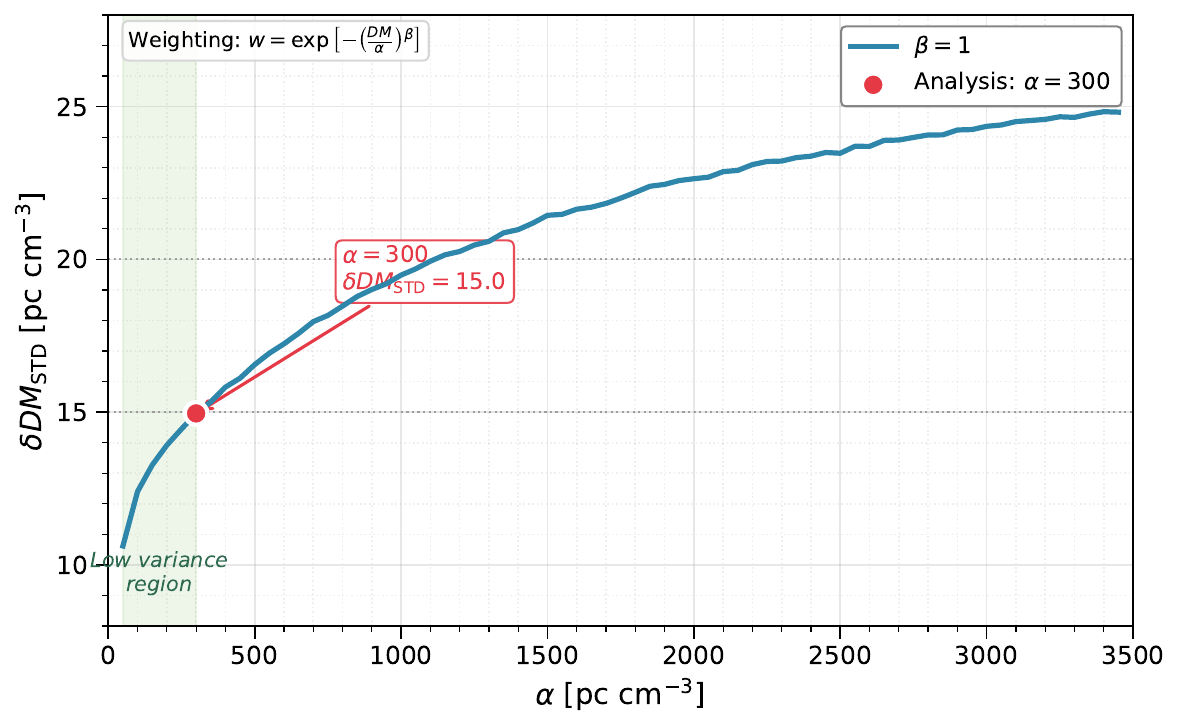}
    \caption{\textbf{Top}: Comparison of the CDFs for the extragalactic DMs \dmeg\
for CHIME/FRBs within 1 virial radius of M31 (magenta) versus the
control sample (blue) for our \mocksampleog\ sample. The AD test yields $p = \adtestdmegmockpvalue$, indicating no significant difference between the two populations at the $p < 0.05$ level. Note the apparent offset at $\dmeg < 800\,\dmunits$ suggesting a non-negligible \dmtarg\ may be retrieved from the lower DM sightlines, which should suffer less contamination from non-M31 contributions. \textbf{Bottom:} Standard deviation of the weighted mean difference (\ddmstandardev) as a function of $\alpha$ at fixed $\beta = 1$. The blue curve shows \ddmstandardev\ computed from 10,000 random draws of 200 FRBs from our mock sample. The green shaded region marks the low-variance regime at small $\alpha$. The red point indicates our fiducial choice of $\alpha = 300\,\dmunits$, which achieves \ddmstandardev $\approx 15\,\dmunits$ while retaining sensitivity to plausible M31 halo DM contributions.}
    \label{fig:mock_ks_test}
    \label{fig:alpha_variance}
\end{figure*}


\begin{figure*}[htbp]
    \centering
    \includegraphics[width=\textwidth]{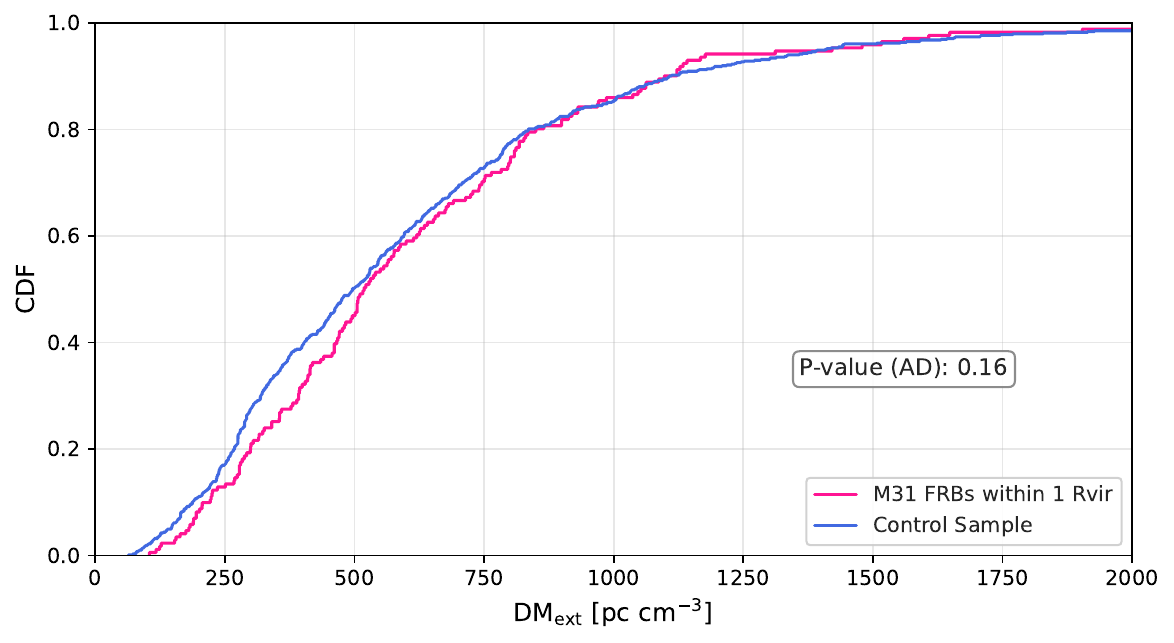}
    \caption{Comparison of the CDFs for the extragalactic DMs \dmeg\
    for CHIME/FRBs within $1\,\rvir$ of M31 (magenta) versus the control
    sample (blue) for the observed data. The AD test yields $p = \adtestdmegpvalue$,
    indicating no significant difference between the two populations at
    the $p < 0.05$ level. Similar to our mock sample
    (Figure~\ref{fig:mock_ks_test}), the apparent offset at
    $\dmeg < 800\,\dmunits$ suggests a non-negligible \dmtarg\ may be
    retrieved from the lower DM sightlines using the weighted estimator.}
    \label{fig:real_ks_test}
\end{figure*}

\section{Results}\label{sec: results}

We now apply the analysis framework developed and validated on mock samples 
(Section~\ref{sec: analysis}) to the observed CHIME/FRB Catalog~2 data \citep{collaboration_second_2026}.

\subsection{Unweighted Comparison}
\label{sec:results_unweighted}

Figure~\ref{fig:real_ks_test} compares the CDFs of \dmeg\ between the M31 and 
control FRB samples in the real data. We find that the two distributions 
are systematically offset for $\dmeg \lesssim 800\,\dmunits$, with M31 sightlines 
exhibiting larger values, before converging at higher DM. Performing
a two-sample Anderson-Darling (AD) test
on these CDFs, we recover an AD test $p$-value of \adtestdmegpvalue.
The test does not detect a statistically significant difference at the
$p < 0.05$ level, consistent with our mock analysis
(Section~\ref{sec:ks_analysis}) which showed that the unweighted CDF
comparison lacks sensitivity to the expected M31 signal.

As in the mock sample, we infer that the signal from M31 is diluted at high 
\dmeg, noting that $\approx 15\%$ of FRBs have $\dmeg > 1000\,\dmunits$. 
This motivates the weighted estimator introduced in 
Section~\ref{sec:weighted mean diff}.

\subsection{Non-Parametric Estimation of the M31 DM Contribution}
\label{sec:results_nonparam}

Following the methodology validated in Section~\ref{sec:nonparam_mock}, 
we apply our non-parametric weighted estimator to the observed sample of 
\numfrbscleansampleinhalo\ M31 FRBs. The FRBs were divided into two radial 
bins corresponding to the inner and outer halo: $0$--$0.5\,\rvir$ and 
$0.5$--$1\,\rvir$, containing \numfrbscontrolsampleinhaloinnerbin\ and 
\numfrbscontrolsampleinhalouterbin\ FRBs, respectively. While finer 
binning is possible in principle, the limited sample size would lead to 
large uncertainties and potentially non-physical (e.g., negative) \ddM\ 
estimates in sparsely populated bins.

The measured values of \ddM\ for each bin are shown in 
Figure~\ref{fig:real_nonparam}. We find $\ddM = \valddmlower$ for the inner 
bin ($0$--$0.5\,\rvir$) with a signal-to-noise ratio of \valddmlowersnr, and 
$\ddM = \valddmupper$ for the outer bin ($0.5$--$1\,\rvir$) with a 
signal-to-noise ratio of \valddmuppersnr. 

In this context, the signal-to-noise ratio is defined as 
${\rm S/N} \equiv \ddM / \sigma_{\rm rand}$, where $\sigma_{\rm rand}$ 
is the standard deviation of $\ddM$ measured along random positions on 
the sky. The random-sky distribution is constructed by repeating the 
same measurement procedure on control sightlines that are not associated 
with M31. The quoted $1\sigma$ uncertainties 
correspond to the dispersion of this Monte Carlo random-sky distribution, 
as described in Section~\ref{sec:nonparam_mock}.

The resulting \ddM\ profile is qualitatively consistent with a 
halo-associated DM contribution from M31. Both bins yield positive \ddM\ 
values that lie above the random-sky expectation (magenta shaded region 
in Figure~\ref{fig:real_nonparam}), with the outer bin showing a stronger 
detection ($\mathrm{SNR} = \valddmuppersnr$). We also show for comparison 
the recent measurements by \citet{Reshma_2025} from two localized FRBs 
with sightlines passing through the inner M31 halo (red diamonds). Their 
estimates of $\ddM \approx 60$--$145\,\dmunits$ at $\rperp \lesssim 0.1\,\rvir$ 
are consistent with our binned measurements and the parametric model fits 
described below.

\subsection{Constraining \texorpdfstring{\rclose}{k\_close} with Parametric Modeling}
\label{sec:results_param}

Our binned measurements of $\ddM(\rperp)$ provide suggestive evidence 
for an M31 CGM detection. We now employ the parametric methodology developed 
in Section~\ref{sec:param_mock} to constrain the closure radius \rclose\ of 
the generalized halo model and derive the corresponding CGM mass.

\begin{figure*}[htbp]
    \centering
    \includegraphics[width=0.9\textwidth]{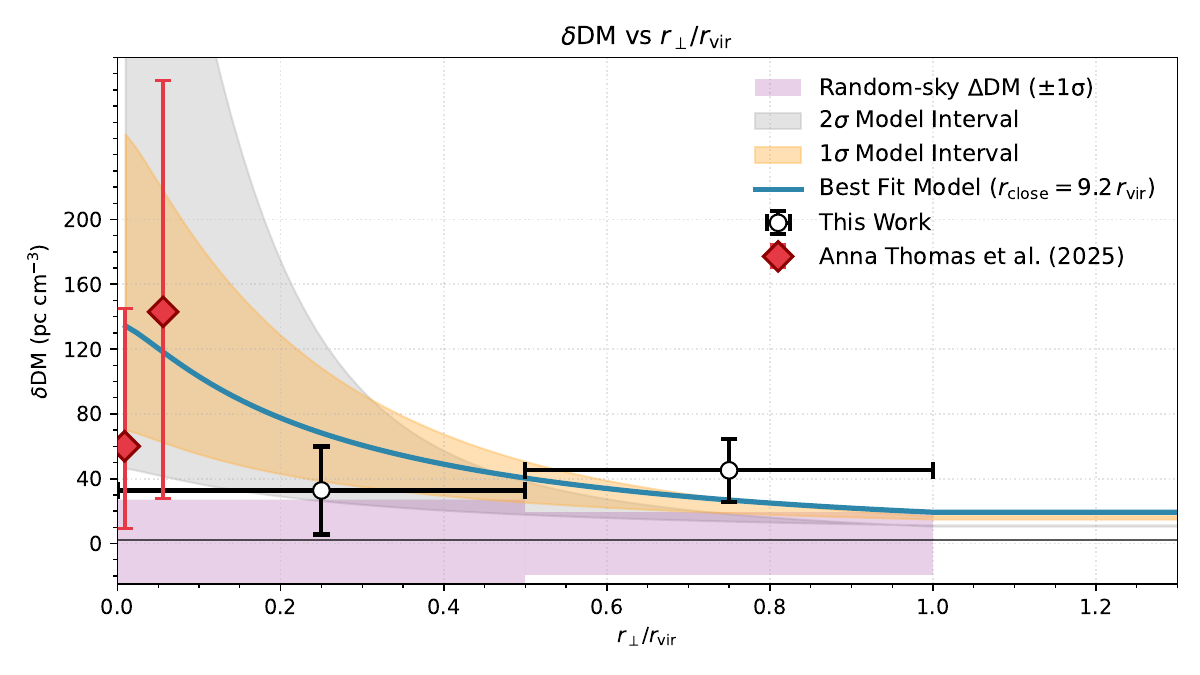}
    \vspace{0.3cm}
    \includegraphics[width=0.9\textwidth]{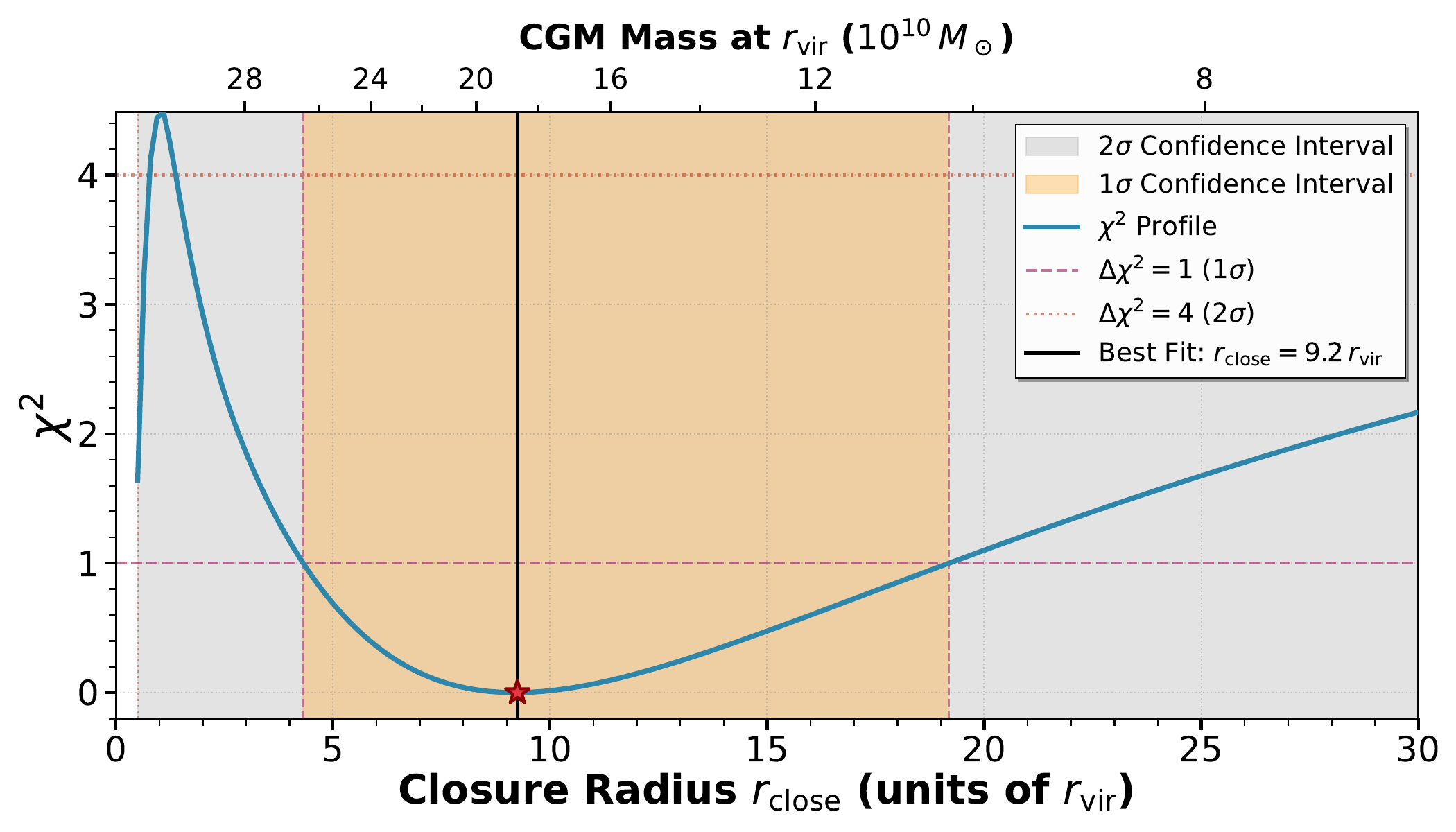}
    \caption{\textbf{Top:} Non-parametric estimation of the M31 halo DM contribution
    from the observed CHIME/FRB Catalog~2 data. Open circles show our
    measurements of \ddM\ in two radial bins ($0$--$0.5\,\rvir$ and
    $0.5$--$1\,\rvir$), with vertical error bars indicating $1\sigma$
    uncertainties and horizontal bars indicating bin widths. Red diamonds
    show independent measurements from \citet{Reshma_2025} based on
    two localized FRBs passing through the inner M31 halo. The magenta
    shaded region shows the $\pm 1\sigma$ range expected for random sky
    positions with no M31 signal. The solid blue curve shows the best-fit
    parametric model ($\kclose = \kclosebestfit\,\rvir$), with orange
    ($1\sigma$) and gray ($2\sigma$) shaded regions indicating model
    uncertainties from the $\chi^2$ analysis (bottom panel).
    \textbf{Bottom:} Parametric constraints on the closure radius \kclose\
    from the observed CHIME/FRB data \citep{collaboration_second_2026}. The blue curve shows the $\chi^2$
    profile as a function of \kclose\ (bottom axis, in units of \rvir).
    The top axis shows the corresponding CGM mass enclosed within \rvir,
    derived from the generalized halo model. Horizontal dashed lines mark
    $\Delta\chi^2 = 1$ and $\Delta\chi^2 = 4$, corresponding to $1\sigma$
    and $2\sigma$ confidence levels. The orange and gray shaded regions
    indicate the corresponding confidence intervals. The red star marks
    the best-fit value $\kclose = \kclosebestfit\,\rvir$, corresponding
    to a CGM mass of $\mcgm = \kclosemassest$.}
    \label{fig:real_nonparam}
    \label{fig:real_chisq}
\end{figure*}

Figure~\ref{fig:real_chisq} shows the $\chi^2$ profile as a function of 
\rclose\ for the observed data. The best-fit value, corresponding to the 
minimum $\chi^2$, is $\rclose = \kclosebestfit\,\rvir$ (red star). The 
$1\sigma$ confidence interval, estimated from $\Delta\chi^2 = 1$, spans 
$\rclose = \kcloseloweronesigma$--$\kcloseupperonesigma\,\rvir$ (orange 
shaded region). The $2\sigma$ interval ($\Delta\chi^2 = 4$) is well 
constrained on the lower-\rclose\ side, yielding 
$\rclose = \kcloselowertwosigma\,\rvir$, which corresponds to a 
$2\sigma$ lower limit on \mcgm. However, toward larger \rclose\ values 
the $\chi^2$ profile flattens within the explored parameter range, and 
thus our analysis does not provide an upper $2\sigma$ bound 
given the adopted limits on \rclose\ (gray shaded region).

The upper axis of Figure~\ref{fig:real_chisq} shows the corresponding 
CGM mass enclosed within \rvir, derived from the generalized halo model 
prescription (Equation~\ref{eqn:Mb_prescription_main}). The best-fit 
closure radius of $\rclose = \kclosebestfit$ corresponds to a CGM mass 
of $\mcgm = \kclosemassest$. This estimate represents the total ionized 
gas mass in the M31 halo within the virial radius, under the assumptions 
of our generalized halo model with full ionization and primordial helium 
abundance.

The best-fit parametric model is overlaid on the non-parametric 
measurements in Figure~\ref{fig:real_nonparam}. The solid blue curve shows 
the model prediction for $\rclose = \kclosebestfit\,\rvir$, while the orange 
and gray shaded regions indicate the $1\sigma$ and $2\sigma$ model 
uncertainties, respectively.

We note that the $\chi^2$ profile is asymmetric, with a steeper rise 
toward small \rclose\ values (concentrated halos) than toward large \rclose\ 
values (extended halos). This reflects the fact that highly concentrated 
models predict large \ddM\ values in the inner halo that are inconsistent 
with our measurements, while extended models with $\rclose \gtrsim 10\,\rvir$ 
produce relatively flat \ddM\ profiles that remain marginally consistent 
with the data. Future observations with larger FRB samples will be needed 
to break this degeneracy and provide tighter constraints on the halo 
structure. Indeed, the $\chi^2$ profile does not reach $\Delta\chi^2 = 4$ on the 
high-\rclose\ side within the explored parameter range, leaving the 
$2\sigma$ upper limit on the closure radius (and correspondingly, the 
$2\sigma$ lower limit on \mcgm) formally unconstrained.
Future observations with larger FRB samples will be needed 
to break this degeneracy and provide tighter constraints on the halo 
structure.


\section{Discussion}\label{sec: Discussion}

In this section, we discuss the implications of our results for the 
baryon content of M31, compare with previous CGM measurements, address 
potential systematic uncertainties, and consider prospects for future 
FRB surveys.

\subsection{Total Mass of Baryons in M31}
\label{sec:baryon_census}

Studies of the CGM indicate that for $L^*$ galaxies with halo masses 
$\sim 10^{12}\,M_\odot$, typically less than 25\% of the cosmic baryonic 
mass \mbcosmic\ of the halo resides in the stellar disk 
\citep[e.g.,][]{Behroozi_2019}. This implies that $\approx 75\%$ of 
\mbcosmic\ is ``missing'' from our view. Either the gas is too diffuse 
for detection with our facilities, or a substantial fraction of baryons 
was expelled beyond the virial radius into the IGM by feedback processes 
\citep[e.g.,][]{bregman2007, mcquinn2016evolution, illya2024}. This inefficiency 
in baryon retention is a key challenge in galaxy formation theory. 
\citet{Kravstov2018} suggest that most galaxies are inefficient at 
converting their available baryons into stars, with Milky Way-mass halos 
typically locking less than 20\% of their expected baryon budget in stars 
and ISM. This inefficiency is shaped by a combination of physical processes, 
including supernova and AGN feedback, galactic winds, and halo-scale heating, 
which prevent baryons from reaching or cooling within the ISM 
\citep[see, for example, the review by][]{hsiao_chen_cgm}. The CGM acts as 
a dynamic reservoir where these processes leave observable imprints.

With this context in mind, we consider a baryonic census of M31. Remarkably, 
assuming a halo mass of $\mhalo = 1.5 \times 10^{12}\,M_\odot$ 
\citep{Van_der_marel_2012}, M31 appears to host a higher-than-typical 
baryonic fraction in stars and inter-stellar Medium (ISM), compared to other $L^*$ systems. This raises the possibility that M31 is unusually efficient at retaining or 
re-accreting its baryons. An independent measurement of the total CGM mass 
is therefore essential to understanding whether M31 is truly atypical or 
simply under-observed in its diffuse gas phases.

\begin{figure*}[htbp]
    \centering
    \includegraphics[width=0.9\textwidth]{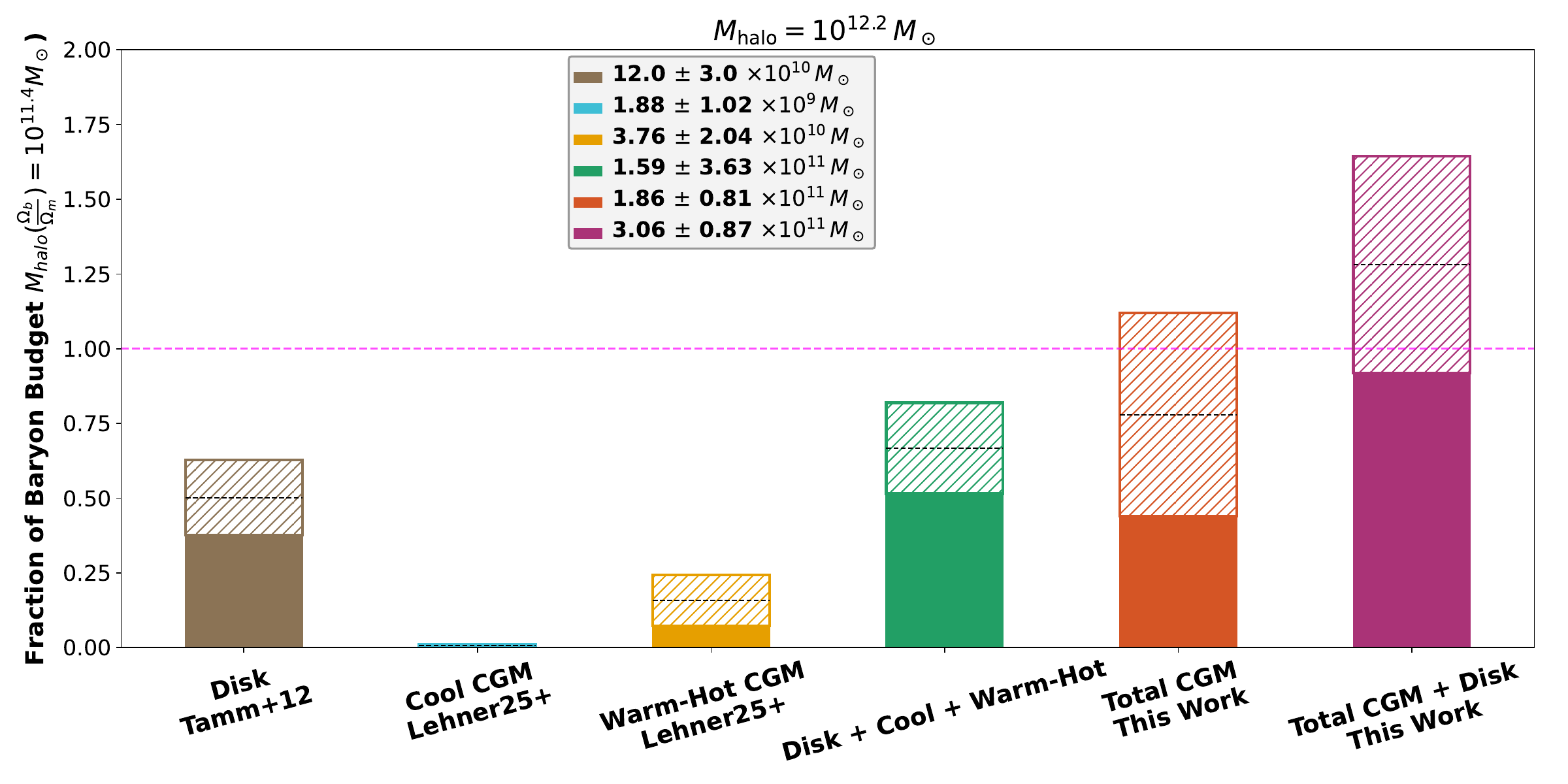}
    \vspace{0.3cm}
    \includegraphics[width=0.7\textwidth]{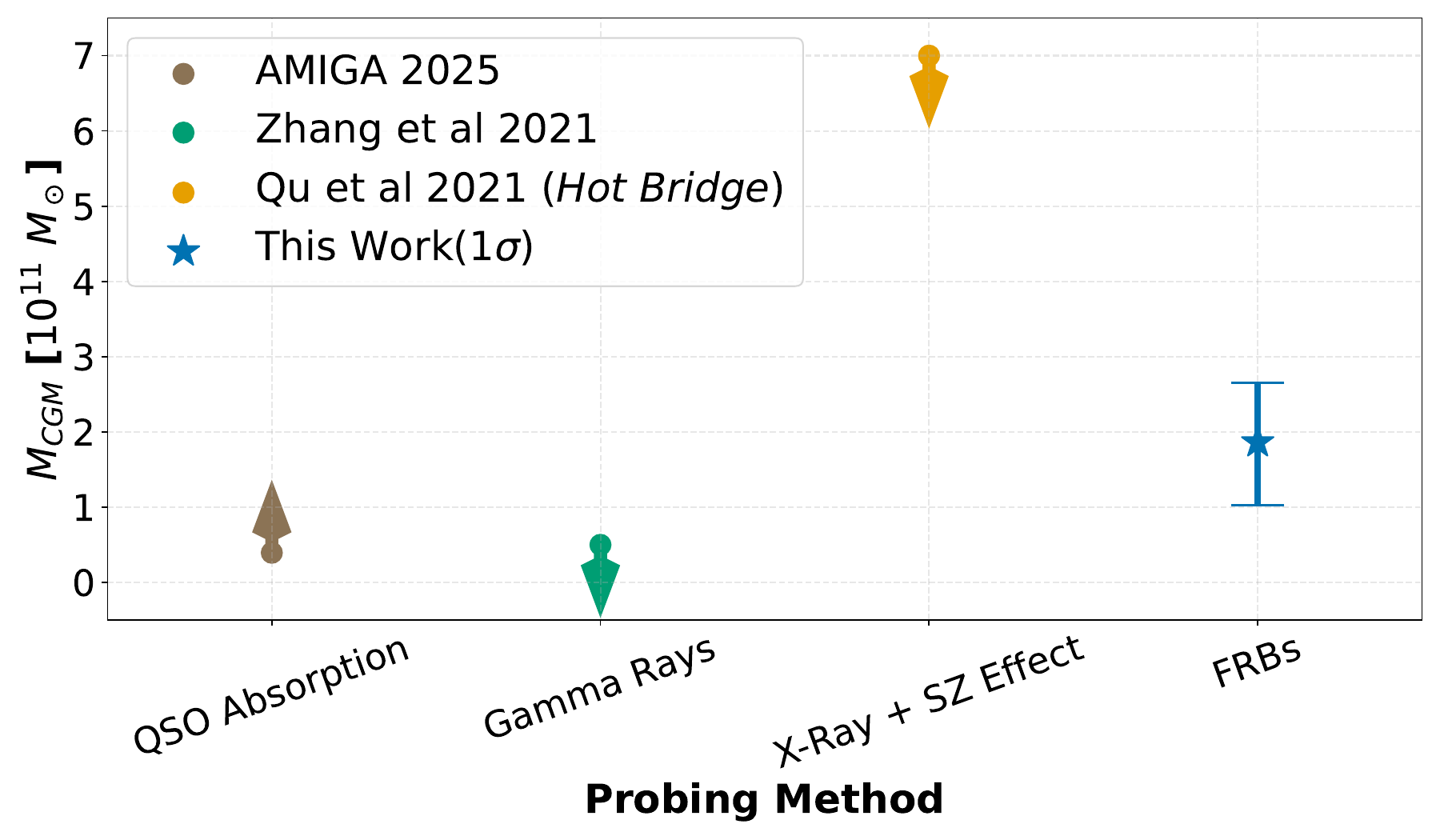}
    \caption{\textbf{Top:} Estimates of the baryonic content in different phases of
    the M31 system, shown as a fraction of the cosmic baryon budget
    $\mbcosmic = \mhalo (\Omega_b/\Omega_m) \approx 10^{11.4}\,M_\odot$.
    From left to right: stellar disk mass from \citet{Tamm_mass_2012}; cool
    CGM mass from \citet{AMIGA_2025}; warm-hot CGM mass from \citet{AMIGA_2025};
    combined disk + cool + warm-hot; total CGM mass from this work; and total
    CGM + disk from this work. Solid bars represent $1\sigma$ lower limits, while
    hatched regions indicate $1\sigma$ uncertainties. The horizontal dashed
    magenta line marks the cosmic baryon fraction.
    \textbf{Bottom:} Comparison of CGM mass estimates for M31 from different
    observational techniques. From left to right: QSO absorption-line
    measurements from AMIGA \citep{AMIGA_2025}; gamma-ray constraints from
    \citet{zhang_gamma_cgm2021}; X-ray + SZ effect upper limit on the ``hot bridge''
    from \citet{Qu2020HotBridge}; and our FRB-based measurement (this work).
    Vertical bars indicate $1\sigma$ uncertainties; arrows indicate upper
    or lower limits. Our FRB measurement of $\mcgm = \kclosemassest$ is
    consistent with the AMIGA lower bounds and the hot-bridge upper limit.}
    \label{fig:baryon_census}
    \label{fig:mcgm_comparison}
\end{figure*}

Adopting $\mhalo = 1.5 \times 10^{12}\,M_\odot$ for the total mass of M31, 
its cosmic baryonic mass corresponds to $\mbcosmic \approx 10^{11.4}\,M_\odot$. 
Figure~\ref{fig:baryon_census} examines estimates for the major baryonic 
components of M31, including our own.  Measurements of the disk by \citet{Tamm_mass_2012} estimate a stellar mass of \stellarmass\ (including the bulge). This 
suggests that the M31 disk corresponds to $\approx 50\%$ of \mbcosmic, notably 
higher than typical $L^*$ galaxies.

Measurements from \citet{AMIGA_2025} infer a mass of 
$\approx 1.88 \pm 1.02 \times 10^{9}\,M_\odot$ and 
$\approx 3.76 \pm 2.04 \times 10^{10}\,M_\odot$ for the cool and warm-hot 
phases of the M31 halo, respectively. It must be noted that the cool and 
warm-hot mass estimates quoted here assume a metallicity of $Z = 0.3\,Z_\odot$ 
for both phases of the CGM, and UV studies are not sensitive to the hot phase.

Our estimate for the total CGM mass from the FRB experiment is 
$\mcgm = \kclosemassest$ (Section~\ref{sec:results_param}). When expressed 
as a fraction of \mbcosmic, this corresponds to $\approx 100^{+22}_{-35}\%$ 
of the expected baryonic content within \rvir. When combined with the stellar 
disk mass, the total baryonic mass would exceed \mbcosmic\ at the $1\sigma$ 
level (Figure~\ref{fig:baryon_census}).

If the combined CGM + disk mass is accurate, this would imply that M31's 
total baryon content approaches or exceeds the fraction predicted by $\Lambda$CDM, 
potentially pointing to ongoing IGM accretion or underestimated systematics in 
our measurement. It supports the growing evidence that the ``missing baryons'' 
in $L^*$ halos may reside in hot, diffuse gas extending to large radii. In 
contrast, simulation results from the SIMBA suite by \citet{illya2024} suggest that
only a small fraction of the ionized baryons are in the CGM, which is in 
agreement with the combined mass of baryons in the cool and warm-hot CGM from 
the UV absorption studies \citep{AMIGA_2025}. Nevertheless, our results 
highlight the potential of FRBs to help close the gap in accounting for the 
diffuse and otherwise difficult-to-detect baryons in the CGM of galaxies.

Figure~\ref{fig:mcgm_comparison} shows our CGM mass estimate in comparison 
with other probes of M31's diffuse gas content. Absorption-line measurements 
toward background quasars \citep{AMIGA_2025} and gamma-ray observations 
\citep{zhang_gamma_cgm2021} yield relatively low estimates of the halo baryon content, 
suggesting a significant baryon deficit. We also illustrate an upper limit 
derived from the hot-bridge structure between the M31 and Milky Way halos, 
inferred from a combination of X-ray and SZ-effect measurements 
\citep{Qu2020HotBridge}. Our FRB-based measurement of 
$\mcgm = \kclosemassest$ lies between these previous estimates: it is 
consistent with the AMIGA lower bounds from \citet{AMIGA_2025} and the 
hot-bridge upper limit from \citet{Qu2020HotBridge}, but is in tension with 
the higher upper bound reported by \citet{zhang_gamma_cgm2021}. This discrepancy may 
be attributed to significant uncertainties in the gamma-ray intensity and 
the adopted cosmic-ray diffusion coefficient in the M31 halo. These results 
demonstrate that FRBs offer a powerful, independent probe of the ionized CGM, 
capable of placing competitive constraints on its total baryonic content.

\subsection{Comparison with Other DM Estimates for Galaxy Halos}
\label{sec:comparison_other_dm}

In our analysis, we estimated \ddM\ to be \valddmlower\ and \valddmupper\ 
with signal-to-noise ratios of \valddmlowersnr\ and \valddmuppersnr\ for the 
two radial bins spanning $0$--$0.5\,\rvir$ (\innerhalorange) and $0.5$--$1\,\rvir$ (\outterhalorange), respectively.

\citet{connor_and_Ravi_halo} used a sample of 474 non-repeating FRBs from the 
CHIME/FRB catalog to estimate the DM contribution of Galactic halos. They 
compared the sightlines of the FRBs with galactic halos $< 40\,\mathrm{Mpc}$ 
from the Gravitational Wave Galaxy Catalog (GWGC). Of the 474 FRBs, 26 
intersected galactic halos at impact parameters ranging over 75--300\,kpc. 
Using the jackknife and Student $t$-test methods without applying any weighting 
schemes, they estimate a DM contribution $> 90\,\dmunits$ at a 95\% confidence 
limit. However, from 12 sightlines that intersect the M31 halo, they found no 
statistically significant DM contribution.

\citet{Wu_McQuinn2023} employed a weighted stacking method on the same sample 
of 474 FRBs. Using a sample of galaxies at $< 80\,\mathrm{Mpc}$, they measure 
a DM contribution ranging over $20$--$50\,\dmunits$ from galactic halos with 
masses $10^{12}$--$10^{13}\,M_\odot$ (similar to M31) to $1\sigma$--$2\sigma$ 
within $0$--$1\,\rvir$ of the galactic centers. 

Most recent measurements by \citet{Reshma_2025} found that the M31 
halo contributes $9$--$145\,\dmunits$ along the sightline of FRB20230903A and 
$28$--$286\,\dmunits$ along the sightline of FRB20230506C. These localized FRB 
measurements probe the inner M31 halo ($\rperp \lesssim 0.1\,\rvir$) and are 
shown in Figure~\ref{fig:real_nonparam} for comparison with our binned estimates. 
The consistency between our statistical analysis and these independent 
single-sightline measurements lends confidence to both approaches.

Our measurements on M31 are consistent with those of \citet{Wu_McQuinn2023} and 
\citet{Reshma_2025}, suggesting that halos of similar masses will have 
similar DM contributions. This serves as evidence of the existence of a 
massive galactic halo around M31, also illustrating the potential of FRBs 
to be used for estimating the masses of galactic halos. With larger samples, 
FRBs can be used to estimate the masses of other nearby halos and structures, 
such as M87 and the Coma Cluster.

\subsection{Systematics}
\label{sec: systematics}

Before concluding, we emphasize several sources of potential systematic 
uncertainty that could affect our estimate of \mcgm.

\textbf{Local Hot Bridge:} One caveat arises from the possible existence of 
a ``Local Hot Bridge,'' a diffuse, X-ray, and SZ-bright plasma structure 
between the Milky Way and M31, first proposed by \citet{Bregman_Lloyd2007}. 
\citet{Qu2020HotBridge} estimate the mass of the hot bridge to range over 
$6 \times 10^{10}\,M_\odot$--$7 \times 10^{11}\,M_\odot$ (see 
Figure~\ref{fig:mcgm_comparison} for a comparison with \mcgm\ measurements 
of M31 by other probes). This feature, inferred through angular geometry and 
modeled as a $\approx 400\,\mathrm{kpc}$ cylindrical filament, may contribute 
significantly to the total electron column density along sightlines toward 
M31. If real, it could add up to $\approx 100$--$400\,\dmunits$ to the 
observed \dmeg, potentially contaminating or even dominating the DM attributed 
to M31's halo. However, the existence and structure of this bridge are derived 
from model-dependent fits to X-ray and SZ residuals and thus remain 
hypothesis-driven. The inferred geometry is consistent with a large-scale 
intergalactic feature, but its contribution remains uncertain without direct 
absorption-line confirmation or off-axis FRB validation. Nonetheless, we 
emphasize that this possibility should be considered when interpreting DM 
measurements near M31, and encourage future work to incorporate multi-wavelength 
tracers to better isolate individual plasma components.

\textbf{Milky Way Halo Contribution:} An additional source of uncertainty in 
our estimates of \dmtarg\ arises from the contribution of the Milky Way's halo. 
\citet{RaviMWhalo23} estimate the \dmmwhalo\ to be $\sim 38 \pm 19\,\dmunits$, 
while \citet{Cook23} report upper limits ranging from 52 to $119\,\dmunits$ 
for high-latitude sightlines ($|b| \geq 30^\circ$). Given these significant 
systematic uncertainties, we elect not to subtract a specific value for the 
Milky Way halo contribution. Instead, we rely on the differential nature of 
our experiment: with over 800 FRBs in our final analysis set, we assume that 
the distribution of \dmmwhalo\ is statistically similar for both M31 and 
control sightlines. This approach allows us to isolate the excess contribution 
from M31 via the weighted mean difference, while acknowledging residual 
uncertainties due to unmodeled variations in the Galactic halo. We note that 
\citet{keating2020exploring} provide a detailed comparison of Milky Way halo 
models that could inform future analyses.

This is a limitation of our current results. In particular, if the Galactic 
halo were asymmetric in its DM contribution above and below the Galactic plane, 
this would confound the current analysis. With larger FRB samples, we may test 
for a radial dependence in the M31 signal from its center on the sky. An 
asymmetric Galactic halo would have to conspire to reproduce any such signature.

\textbf{CHIME Sensitivity Variations:} Because CHIME's sensitivity varies 
across the sky, FRBs intersecting the M31 halo could be detected more 
efficiently than those in other regions, and therefore bias M31 FRBs towards 
higher DM. This introduces a potential systematic when comparing M31 FRBs and 
the control sample FRBs. We address this by incorporating a weighting scheme 
(Section~\ref{sec:weighted mean diff}) that accounts for differences in the 
\dmeg\ distributions, which are affected by sensitivity variation. Although 
CHIME is more sensitive toward declinations near M31, which could skew the 
M31-intersecting sample toward higher-DM FRBs, this weighting scheme mitigates 
that potential bias by reducing the influence of high-DM outliers on the 
inferred M31 signal.

\subsection{Future Surveys}
\label{sec:future}

In our analysis, we have used a sample of arcminute-localized FRBs to estimate 
the \ddM\ contributions of M31. We divided the halo into two radial bins and 
measured \ddM\ with signal-to-noise ratios of \valddmlowersnr\ and 
\valddmuppersnr. Our methodology supports further binning of the halo, enabling 
a more refined estimation of the \ddM\ profile. However, increasing the number 
of bins with the current sample reduces the signal-to-noise ratio of \ddM\ 
measured in each bin. To explore the potential of this approach in future 
surveys, we applied our methodology to simulated FRB samples of varying sizes. 
We found that for a sample of 20,000 FRBs (approximately 1,000 intersecting the 
M31 halo), with a sky distribution similar to the CHIME/FRB Catalog~2 sample \citep{collaboration_second_2026}, 
a \ddM\ profile can be recovered with $> 90\%$ confidence using up to five bins.

Achieving the sample size of 20,000 FRBs that we considered in simulations will
require sustained observations over the coming years. In the near term, CHIME/FRB
is undergoing upgrades expected to increase its detection rate by a factor of
$\sim 3$, potentially yielding $\sim 6$ FRBs per day. At this rate, accumulating
20,000 FRBs would require approximately 5--6 years of continued operation,
making CHIME a viable path to the sample sizes considered in our simulations.

Looking further ahead, next-generation wide-field surveys such as the Canadian
Hydrogen Observatory and Radio-transient Detector (CHORD), the Bustling Universe
Radio Survey Telescope in Taiwan (BURSTT), and the Deep Synoptic Array 2000
(DSA-2000) will dramatically increase global FRB detection rates. CHORD, expected
to be fully operational by 2027, will provide broader frequency coverage and
significantly higher sensitivity than CHIME, with forecasts indicating FRB
detection rates up to a factor of 3--6 greater than CHIME/FRB's current
$\sim 2$ per day (CHORD in prep). DSA-2000, scheduled to begin operations in
2028, is expected to detect over 25,000 FRBs per year. However, as general-purpose
survey instruments, neither CHORD nor DSA-2000 will be continuously observing the
M31 region, so the fraction of their total FRB yield that probes M31 sightlines
will depend on their sky coverage and survey strategy.

BURSTT, which began operations in mid-2025, features a vast field of view and is
expected to double the detection rates.
Importantly, its sensitivity to low-DM bursts makes it particularly valuable for
probing nearby FRBs \citep{Ling2023}, where contributions from halos such as M31
are more prominent, thereby enhancing the measurable \ddM\ signal.

Collectively, these instruments will rapidly expand the available FRB sample,
though the rate at which M31-intersecting sightlines accumulate will depend not
only on total detection rates but also on the sky coverage overlap with the M31
direction.

Beyond sample size, future surveys will provide improved FRB localizations 
and, in many cases, host galaxy identifications with spectroscopic redshifts. 
Instruments such as ASKAP and MeerTRAP already achieve sub-arcsecond localizations, 
enabling direct host associations and precise impact parameter measurements. For 
M31 specifically, localized FRBs with known redshifts would allow subtraction of 
the \dmcosmic\ and \dmhost\ contributions, isolating \dmtarg\ on a 
sightline-by-sightline basis rather than statistically. The recent work by 
\citet{Reshma_2025} demonstrates the power of this approach, achieving 
constraints on individual M31 sightlines that complement our statistical analysis.

Taken together, these facilities suggest that recovering the detailed \ddM\ 
profile of M31 is well within reach: CHORD and DSA-2000 will deliver the large 
samples necessary, while BURSTT's emphasis on nearby events offers a potent 
probe of the M31 halo. Combined with improved localizations from ASKAP 
and MeerTRAP, the next decade of FRB observations promises to transform our 
understanding of the CGM in M31 and other nearby galaxies.


\section{Conclusion}\label{sec: Conclusion}

This work has leveraged FRBs to statistically constrain the baryonic mass 
of M31's CGM, a key but elusive reservoir that regulates galaxy evolution 
and baryon cycling. Traditional methods for probing the CGM face severe 
limitations, especially for nearby halos like M31. Our analysis demonstrates 
that FRBs offer a powerful, redshift-independent alternative.

Using a control sample of \numfrbscontrolsample\ FRBs and 
\numfrbscleansampleinhalo\ M31 FRBs, we estimated the halo's contribution 
to the dispersion measure (DM) by comparing to a control sample with similar 
Galactic DM. We constrained the M31 DM contribution, \ddM, to \valddmlower\ 
and \valddmupper\ in radial bins of impact parameters spanning 
\innerhalorange\ and \outterhalorange, respectively. 
We then used the parametric $\chi^2$ framework with our generalized 
halo model to constrain the closure radius to 
$\kclose = \kclosebestfit^{+\kcloseupperonesigma}_{-\kcloseloweronesigma}\,\rvir$ 
($1\sigma$), corresponding to a total CGM mass of $\mcgm = \kclosemassest$.

\begin{itemize}
    \item Our measurements provide the first direct constraint on M31's CGM mass 
    using FRBs, and suggest the presence of a substantial ionized gas reservoir.
    
    \item While \ddM\ estimates enable characterization of the baryon density 
    profile, current uncertainties are still substantial in both radial bins, limiting 
    our ability to resolve the profile in detail. We estimate that 
    $\sim$20,000 FRBs will be needed to constrain the radial structure of 
    the CGM to a confidence limit $> 90\%$.
    
    \item Our analysis is based on a technique that requires only coarse FRB 
    localizations and no host galaxy redshift information, making it scalable 
    to future surveys.
\end{itemize}

Looking forward, as CHIME/FRB, ASKAP, MeerTRAP, and upcoming experiments like 
CHORD, BURSTT, and DSA-2000 expand the FRB sample size, we anticipate transformative 
improvements in CGM studies. Our method can be readily applied to other 
massive nearby halos, enabling a statistical inventory of circumgalactic 
baryons in the local universe.

These results represent a step toward establishing FRBs as precision tools 
for mapping diffuse baryons and unlocking the role of galaxy halos in the 
cosmic matter cycle.

\FloatBarrier

\section{Acknowledgments}
Authors L.K., J.X.P., and S.S., as members of the Fast and Fortunate for FRB Follow-up team, acknowledge support from NSF grants AST-1911140 and AST-1910471. C. L. acknowledges support from the Miller Institute for Basic Research at UC Berkeley. A.B.P. acknowledges support by NASA through the NASA Hubble Fellowship grant HST-HF2-51584.001-A awarded by the Space Telescope Science Institute, which is operated by the Association of Universities for Research in Astronomy, Inc., under NASA contract NAS5-26555. A.B.P. also acknowledges prior support from a Banting Fellowship, a McGill Space Institute~(MSI) Fellowship, and a Fonds de Recherche du Quebec -- Nature et Technologies~(FRQNT) Postdoctoral Fellowship. K.W.M. is supported by NSF Grant Nos. 2008031, 2510771 and holds the Adam J. Burgasser Chair in Astrophysics. E.F. is supported by the National Science Foundation (NSF) under grant number AST-2407399. The authors acknowledge helpful discussions with Mathew McQuinn, Isabel Medlock, Gwendolyn Eadie, and Raja Guhathakurta. 

We acknowledge that CHIME and the \kkoname\ Outrigger (KKO) are built on the traditional, ancestral, and unceded territory of the Syilx Okanagan people. \kkoname\ is situated on land leased from the Imperial Metals Corporation. We are grateful to the staff of the Dominion Radio Astrophysical Observatory, which the National Research Council of Canada operates. CHIME operations are funded by a grant from the NSERC Alliance Program and by support from McGill University, the University of British Columbia, and the University of Toronto. CHIME was funded by a grant from the Canada Foundation for Innovation (CFI) 2012 Leading Edge Fund (Project 31170) and by contributions from the provinces of British Columbia, Qu\'ebec, and Ontario. The CHIME/FRB Project was funded by a grant from the CFI 2015 Innovation Fund (Project 33213) and by contributions from the provinces of British Columbia and Qu\'ebec, and by the Dunlap Institute for Astronomy and Astrophysics at the University of Toronto. Additional support was provided by the Canadian Institute for Advanced Research (CIFAR), the Trottier Space Institute at McGill University, and the University of British Columbia. The CHIME/FRB baseband recording system is funded in part by a CFI John R. Evans Leaders Fund award to IHS.

Software: astropy \citep{astropy:2022}, pandas \citep{reback2020pandas}, and numpy \citep{harris2020array}

\FloatBarrier

\newpage
\appendix


\FloatBarrier
\section{Sensitivity to Weighting Parameter \texorpdfstring{$\alpha$}{alpha}}\label{app:alpha_sensitivity}

In Section~\ref{sec:weighted mean diff}, we adopted a fiducial weighting 
parameter of $\alpha = 300\,\dmunits$ with $\beta = 1$ for our weighted 
estimator (Equation~\ref{eqn:weight}). Here we demonstrate that our main 
results are not sensitive to this choice by repeating our analysis for 
$\alpha = 150, 200, 400, 500$, and $600\,\dmunits$.

\begin{figure*}[htbp]
    \centering
    \includegraphics[width=0.48\textwidth]{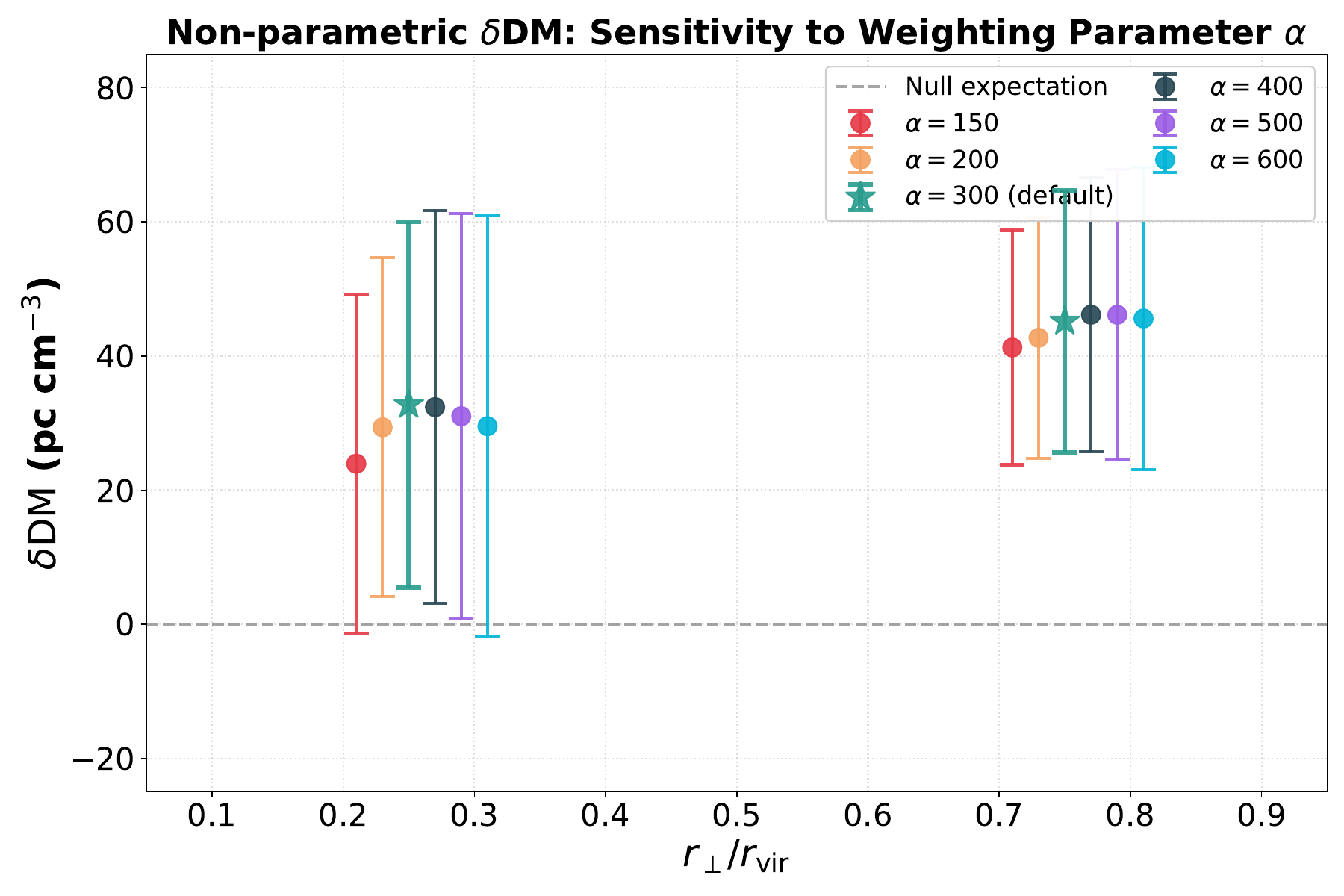}
    \hfill
    \includegraphics[width=0.48\textwidth]{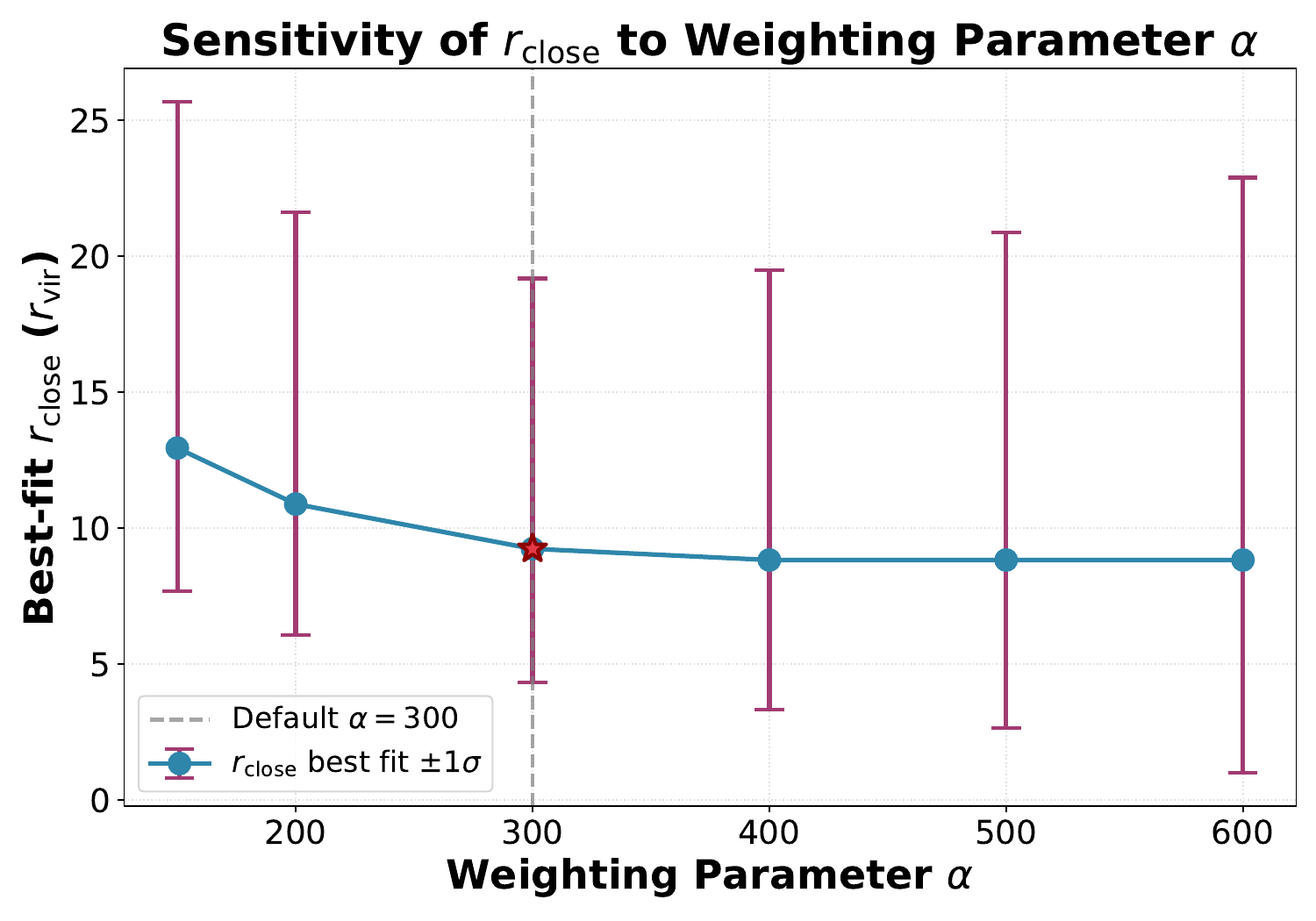}
    \caption{\textbf{Left:} Sensitivity of the non-parametric \ddM\ estimates to the
    weighting parameter $\alpha$. Points show the measured \ddM\ in the inner
    (left, $0$--$0.5\,\rvir$) and outer (right, $0.5$--$1\,\rvir$) radial bins
    for $\alpha = 150, 200, 300, 400, 500$, and $600\,\dmunits$. The fiducial
    value $\alpha = 300$ is marked with a star. Error bars indicate $1\sigma$
    uncertainties. The horizontal dashed line shows the null expectation ($\ddM = 0$).
    Note that the horizontal offsets are for clarity, each error bar represents measurements for either the inner bin or the outer bin.
    \textbf{Right:} Sensitivity of the best-fit closure radius \kclose\ to $\alpha$.
    Blue points show the best-fit \kclose\ values with $1\sigma$ error bars.
    The vertical dashed line marks $\alpha = 300$, with the best-fit shown as a red star.
    Both panels demonstrate that our conclusions are robust to the choice of $\alpha$.}
    \label{fig:alpha_ddm}
    \label{fig:alpha_kclose}
\end{figure*}

The left panel of Figure~\ref{fig:alpha_ddm} shows the non-parametric \ddM\ estimates 
in the inner (\innerhalorange) and outer (\outterhalorange) radial bins 
as a function of $\alpha$. The measured \ddM\ values are consistent across 
all choices of $\alpha$, with the fiducial $\alpha = 300$ (star markers) 
lying well within the range spanned by other values. The error bars increase 
for smaller $\alpha$ values, reflecting the reduced effective sample size 
when more aggressive downweighting is applied.

The right panel of Figure~\ref{fig:alpha_kclose} shows the best-fit closure radius \kclose\
as a function of $\alpha$. The best-fit values range over
$\kclose \approx 9$--$13\,\rvir$ across the explored range, with all values
consistent within the $1\sigma$ uncertainties. The fiducial $\alpha = 300$
yields $\kclose \approx 10\,\rvir$, near the middle of this range. The
$1\sigma$ confidence intervals overlap substantially for all $\alpha$ values,
confirming that the choice of weighting parameter does not significantly
affect our conclusions.

We conclude that our results are robust to the choice of $\alpha$ within 
the range $150$--$600\,\dmunits$. The fiducial value $\alpha = 300\,\dmunits$ 
lies within the optimal region identified in Figure~\ref{fig:alpha_variance} 
and provides a balance between variance reduction and signal retention.

\FloatBarrier
\newpage
\bibliographystyle{aasjournal}
\bibliography{references}

\allauthors


\end{document}